\documentclass[a4paper,11pt]{article}
 \pdfoutput=1
 \usepackage{lscape}
 \usepackage[all]{xy} 
 \usepackage{longtable} 
 \usepackage{jheppub} 
 \usepackage{graphicx}
 \usepackage{amssymb}
 \usepackage{amsmath}
\usepackage{mathtools}
\usepackage{listings}
\usepackage{dcolumn}
\usepackage{bm}
\usepackage{color}
\usepackage{multirow}
\usepackage{upquote}

\allowdisplaybreaks

 \newcommand{\lsim}{{\;\raise0.3ex\hbox{$<$\kern-0.75em\raise-1.1ex\hbox{$\sim$}}\;}}
\newcommand{\gsim}{{\;\raise0.3ex\hbox{$>$\kern-0.75em\raise-1.1ex\hbox{$\sim$}}\;}}
\newcommand{\beq}{\begin{equation}}
\newcommand{\eeq}{\end{equation}}
\newcommand{\bea}{\begin{eqnarray}}
\newcommand{\eea}{\end{eqnarray}}
\mathchardef\minus="002D

\title{\boldmath The Hierarchy Solution to the LHC Inverse Problem}
\author[a]{James~S.~Gainer\note{Corresponding author: {\tt
      jgainer@ufl.edu}},}
\author[a]{Konstantin T.~Matchev,}  
\author[b,c,d]{Myeonghun Park\note{Corresponding author: {\tt
      parc@apctp.org}},}

\affiliation[a]{Physics Department, University of Florida, Gainesville, FL 32611, USA.}
\affiliation[b]{Asia Pacific Center for Theoretical Physics, San 31, Hyoja-dong, Nam-gu, Pohang 790-784, Korea} 
\affiliation[c]{Department of Physics, Postech, Pohang 790-784, Korea} 
\affiliation[d]{Kavli IPMU (WPI), The University of Tokyo, Kashiwa, Chiba 277-8583, Japan}

\abstract{
Supersymmetric (SUSY) models, even those described by relatively few
parameters, generically allow many possible SUSY particle (sparticle)
mass hierarchies.  As the sparticle mass hierarchy determines, to a
great extent, the collider phenomenology of a model, the enumeration
of these hierarchies is of the utmost importance.  We therefore
provide a readily generalizable procedure for determining
the number of sparticle mass hierarchies in a given SUSY model. As an
application, we analyze the gravity-mediated SUSY breaking scenario
with various combinations of GUT-scale boundary conditions involving
different levels of universality among the gaugino and scalar
masses. For each of the eight considered models, we provide the
complete list of forbidden hierarchies in a compact form.  Our main
result is that the complete (typically rather large) set of forbidden
hierarchies among the eight sparticles considered in this analysis can
be fully specified by just a few forbidden relations involving much
smaller subsets of sparticles.}

\date{April 14, 2015}
\preprint{
\begin{flushright} 
IPMU14-0305
\\
APCTP Pre2015 - 009
\end{flushright} 
}

\begin{document} 
\maketitle
\flushbottom

\section{Introduction}
\label{sec:introduction}

Low-energy supersymmetry (SUSY)~\cite{Martin:1997ns} is a
well-motivated paradigm for new physics beyond the Standard Model, as it
explains the lightness of the recently-discovered Higgs
boson~\cite{Aad:2012tfa, Chatrchyan:2012ufa}, provides a dark matter
candidate~\cite{Jungman:1995df}, and, in some models, allows for the
unification of gauge couplings, consistent with the low energy
data~\cite{Feng:2013pwa}.  That there is currently no compelling
evidence of sparticle production at the CERN Large Hadron Collider (LHC)
\cite{SUSYATLAS, SUSYCMS}
strongly motivates efforts to make sure that ``no stone is left
unturned" in searching for SUSY and/or setting more rigorous limits on
the masses of sparticles.  It is therefore important to understand all
possible signatures of SUSY models; the rich phenomenology of SUSY
means that there are unexplored ``corners" of even well-known models
like the CMSSM~\cite{Cohen:2013kna}. Such studies will become
particularly valuable once a signal of new physics is observed, as
they will help determine whether we are seeing a SUSY signal in the
first place, and if so, what variety of SUSY model is responsible for
it.

However, the exploration of SUSY collider phenomenology in full
generality is a very challenging endeavor, complicated by the
multitude of scenarios and the large number of SUSY parameters. 
Several approaches have been tried:
\begin{itemize}
\item {\em Analysis of specific benchmark points}~\cite{Hinchliffe:1996iu,
    Abdullin:1998pm, Battaglia:2001zp, Allanach:2002nj, De
    Roeck:2005bw, AbdusSalam:2011fc}.  The advantage of this approach
  is that it relies on well motivated and popular theory models.
  Benchmark points provide concrete, clear targets for discovery, and
  allow the sensitivities of different types of experiments to be
  compared in a meaningful way~\cite{Ellis:2001hv, Eigen:2001mk,
    Arrenberg:2013rzp}.  The downside is that the conclusions are very
  model-dependent and cannot be easily generalized to any arbitrary
  SUSY scenario.
\item {\em Analysis of the (phenomenologically relevant) MSSM parameter
  space (pMSSM)}~\cite{Berger:2008cq, Cotta:2009zu, AbdusSalam:2009qd,
    Sekmen:2011cz, Arbey:2011un, CMS:2013rda}.  The advantage here is
  that in principle one is exploring all corners of parameter space
  and encountering all phenomenologically interesting signatures.  The
  disadvantage is that the large dimensionality of the pMSSM parameter
  space makes full coverage impossible. This has prompted studies
  in a correspondingly smaller parameter subspace, e.g.,~a nine
  parameter subset (pMSSM9)~\cite{Fowlie:2013oua}.
  Alternatively, one could restrict one's attention to a specific
  sector of the model, e.g.,~the four lightest states in the new
  physics particle spectrum ~\cite{Feldman:2007zn, Feldman:2008hs,
    Berger:2008cq}.  (See also the simplified model approach below.)
\item {\em Analysis of simplified models.}  The third, intermediate,
  approach is motivated by signature-based searches for new
  physics at the LHC.  For any given experimental signature, one may
  consider only the particles that are relevant to this  specific
  channel, arriving at a so-called ``simplified model" with only a
  handful of parameters~\cite{Alwall:2008ag, Alves:2011wf}. This
  approach inherits, to a degree, the advantages of the previous two
  methods, but the connection to the underlying high energy theory
  (and its fundamental parameters) is obscured.
\end{itemize}

The difficulty in exploring the phenomenology of the pMSSM parameter
space in full generality manifests itself in the  ``LHC inverse
problem"~\cite{ArkaniHamed:2005px} of mapping an experimentally
observed set of signatures at the LHC to
a  specific parameter point in theory space. The main challenge stems
from the vastness of the SUSY parameter space --- even under ideal
circumstances the map is not unique and exhibits
degeneracies. Furthermore, such maps so far have been constructed by
(relatively sparse) scans in parameter space~\cite{Berger:2007yu,
  Altunkaynak:2008ry, Kneur:2008ur, Balazs:2009it}, and it is not
clear to what extent the derived conclusions are robust and reliable.

A possible resolution to these problems was put forward in
Ref.~\cite{Konar:2010bi}, which proposed a more manageable and
practical parameterization of theory space.  The main idea is that the
generic parameter space $\mathbb{R}^n$ of $n$ SUSY parameters can be thought of
as the direct product of the set of all possible permutations $S_n$ of
those $n$ parameters and the remaining coset $\mathbb{R}^n/S_n$:
\beq
\mathbb{R}^n = S_n \otimes \mathbb{R}^n/S_n.
\label{decomposition}
\eeq
The $n$ parameters considered in Ref.~\cite{Konar:2010bi} were all SUSY
mass parameters, hence the permutations in $S_n$ were named ``mass
hierarchies", or ``hierarchies" for short.  The advantage of the
decomposition (\ref{decomposition}) becomes evident when we consider
the nature of a new physics signal at a collider and, in particular,
the dependence on the two factors $S_n$ and $\mathbb{R}^n/S_n$. In general, the
collider phenomenology of any new physics model depends on certain
{\em qualitative} and {\em quantitative} aspects, where the former are
parameterized by $S_n$, while the latter depend mostly on $\mathbb{R}^n/S_n$:
\begin{itemize}
\item {\em Quantitative aspects.} Those are the factors which
  determine the overall signal rate, such as: 
\begin{enumerate}
\item {\em Signal production cross section.} This depends on the
  magnitudes of the sparticle {\em masses}, which are encoded in the
  $\mathbb{R}^n/S_n$ factor alone.  Generally, heavier particles have smaller
  cross-sections and vice versa.
\item {\em Signal branching fractions.} These are functions of the
  magnitudes of the {\em mass splittings}, which are also part of
  $\mathbb{R}^n/S_n$. For example, decay modes that are close to threshold, are
  kinematically suppressed, and the amount of suppression depends on
  exactly how close to threshold we are, i.e.,~on the size of the mass
  splitting.
\item {\em Signal acceptances and efficiencies.} These depend on the
  hardness of the SM decay products observed in the detector, which in
  turn is again a function of the magnitudes of the {\em mass
    splittings}. Parameter space points in $\mathbb{R}^n/S_n$ with large mass
  splittings lead to harder leptons, jets, etc., and correspondingly
  higher efficiencies. On the other hand, smaller mass splittings lead
  to degenerate scenarios with lower efficiencies, where discovery
  becomes problematic, see e.g.~\cite{Alves:2011sq, LeCompte:2011cn,
    Alves:2012ft}.
\item {\em Relative contribution of strong versus electroweak
    production.}  In the years leading up to the LHC, it was usually
  assumed that strong SUSY production would dominate, and the first
  sign of SUSY would most likely be seen in squark and/or gluino
  production. Such expectations are based on the fact that strong
  production is enhanced due to the large strong coupling and the
  color multiplicity factors.  However, this expectation is not true
  in models with heavy colored superpartners and light electroweak
  superpartners, where strong production is kinematically suppressed
  relative to electroweak production. The amount of suppression
  depends on the relative size of the superpartner masses, which is
  again parameterized by $\mathbb{R}^n/S_n$.
\end{enumerate}
\item {\em Qualitative aspects.} These are the factors that determine
  the {\em type} of discovery signature one is looking for, namely,
  the {\em identity} and {\em multiplicity} of SM particles in the
  final state. These qualitative features are mostly determined by the
  mass {\em hierarchy}, and can be parameterized by an element of the
  $S_n$ factor in (\ref{decomposition}).
\end{itemize}

The crucial observation of Ref.~\cite{Konar:2010bi} was that, as long
as we are interested in the {\em qualitative} aspects of SUSY collider
phenomenology, we are justified in focusing on $S_n$ only and studying
hierarchies of sparticles without any reference to the actual sizes of
their masses. 
Note that the number of elements in $S_n$ is a finite number (namely,
$n!$), which allows a fully exhaustive exploration and classification
of the sets of experimental signatures associated with each hierarchy.
Somewhat surprisingly, such studies were able to reveal previously
overlooked corners of pMSSM parameter space with dramatic, yet
relatively unexplored, multi-lepton signatures~\cite{Konar:2010bi}.
While the analysis of~\cite{Konar:2010bi} was originally applied only
to pMSSM9, it was subsequently extended to include third generation
sfermions and/or $R$-parity violation in the MSSM
~\cite{Dreiner:2012wm} and to the NMSSM~\cite{Dreiner:2012ec}.

In this paper we build on the work in \cite{Konar:2010bi,
  Dreiner:2012wm, Dreiner:2012ec} and demonstrate the 
cataloguing of all possible hierarchies within specific SUSY breaking
models.
Following~\cite{Konar:2010bi}, we shall consider only the sparticles
shown in table~\ref{tab:summary}. Their mass spectrum is parameterized
by the parameters listed in the last row of the table. Some of these
parameters correspond to several particles; for instance, the wino
mass $M_W$ describes the mass of both a mostly wino-like neutralino
$\tilde{w}^0$ and a mostly wino-like chargino $\tilde{w}^\pm$.  (The
remaining two gaugino mass parameters are denoted as $M_B$, the bino
mass, and $M_G$, the gluino mass.) Likewise, we consider two
degenerate generations of sfermions, so that each of the five sfermion
masses ($M_Q$, $M_U$, $M_D$, $M_L$ and $M_E$) describes particles in
both the first and second generation.  Finally, in the case of
left-handed squarks (sleptons) the parameter $M_{Q}$ ($M_L$) refers to
both members of the isodoublet, $\tilde{u}_L$ and $\tilde{d}_L$
($\tilde{e}_L$ and $\tilde{\nu}_L$).  Tree-level sparticle mixing and
one-loop corrections~\cite{Pierce:1996zz} complicate this story
somewhat, since the experimentally measured mass eigenvalues are
slightly different from the soft mass parameters in
table~\ref{tab:summary}. In what follows, we shall assume that the
soft mass parameters 
\beq
\left\{ M_Q, M_U, M_D, M_L, M_E, M_B, M_W, M_G  \right\}
\label{eq:measured}
\eeq
have been extracted from the data via a global fit along the lines
of Refs.~\cite{Desch:2003vw, Lafaye:2004cn, Bechtle:2004pc, Boehm:2013qva}.

\begin{table}[t]
\centering
\begin{tabular}{| c | c | c | c | c | c | c | c | c |}
\hline
 $\tilde{u}_L$,  $\tilde{d}_L$ & $\tilde{u}_R$ & $\tilde{d}_R$ &
 $\tilde{e}_L$,  $\tilde{\nu}_L$ & $\tilde{e}_R$ 
 & $\tilde{g}$ & $\tilde{w}^\pm$,$\tilde{w}^0$ & $\tilde{b}^0$ \\ \hline\hline
$Q$ & $U$ & $D$ & $L$ & $E$ & $G$ & $W$ & $B$ \\ \hline
$1$ & $2$ & $3$ & $4$ & $5$ & $6$ & $7$ & $8$ \\ \hline
$M_Q$ & $M_U$ & $M_D$ & $M_L$ & $M_E$ & $M_G$ & $M_W$ & $M_B$  \\ \hline
\end{tabular}
\caption{\label{tab:summary}
The set of SUSY particles considered in this analysis, shorthand
notation (symbolic and numeric) for each multiplet, 
and the corresponding soft SUSY breaking mass parameter.  
Note our slightly unconventional notation for the three
gaugino masses: $\left\{M_B, M_W, M_G\right\}$ 
instead of $\left\{M_1,  M_2, M_3\right\}$ --- in this paper the latter
are reserved for the corresponding boundary conditions 
(\ref{M1def}-\ref{M3def})
at the GUT scale. }
\end{table}

Given the eight experimentally measured mass parameters of
table~\ref{tab:summary}, the main question which we address in this
paper is the following: {\em can these measurements, by themselves,
  rule out specific SUSY-breaking scenarios?}  For our
purposes here, a SUSY-breaking scenario is nothing more than a set of
boundary conditions for the soft mass parameters of table
\ref{tab:summary} at some high energy scale.  For concreteness, we
shall illustrate our method with a SUGRA-inspired scenario~\cite
{Chamseddine:1982jx, Barbieri:1982eh, Hall:1983iz} in which the
boundary conditions are specified at the GUT scale.  We shall consider
several popular variations of this SUGRA model, in which one imposes
different sets of boundary conditions:
\begin{enumerate}
\item {\em Scalar mass unification into $SU(5)$ multiplets.} Since the
  $SU(5)$ is unbroken above the GUT scale $M_{GUT}$, we shall always
  assume that squarks and sleptons belonging to the same $SU(5)$
  multiplet have a common mass at $M_{GUT}$:
\beq
M_Q(M_{GUT}) = M_U(M_{GUT}) = M_E(M_{GUT}) \equiv M_{10}, \quad
M_L(M_{GUT}) = M_D(M_{GUT}) \equiv M_{5}.
\label{SU5}
\eeq 
In general, $M_{5}$ and $M_{10}$ are independent input parameters ---
even if they start out equal at the Planck scale, they will be
separated due to RGE running from the Planck scale down to the GUT
scale~\cite{Polonsky:1994sr, Polonsky:1994rz} or due to $D$-term
contributions \cite{Kolda:1995iw}.
\item {\em Scalar mass unification into $SO(10)$ multiplets.}
A more constrained version of the model arises if we assume
$SO(10)$-like unification, where
\beq
M_5 = M_{10} \equiv M_{0}.
\label{SO10}
\eeq
\item {\em Gaugino mass unification.} In $SU(5)$ GUTs, the gaugino masses may exhibit
  (some combination of) up to four\footnote{One for each irreducible representation appearing
  in the symmetric product $(24\times 24)_s$ of two adjoints in $SU(5)$.} different patterns
 \cite{Anderson:1996bg, Anderson:1999uia}, so that in general
\bea
 M_{B}(M_{GUT}) &\equiv & M_1, \label{M1def} \\ [2mm]
 M_{W}(M_{GUT}) &\equiv& M_2, \label{M2def} \\ [2mm] 
 M_{G}(M_{GUT}) &\equiv& M_3 \label{M3def}
\eea can be taken
  as free parameters at the GUT scale.  We shall also optionally
  consider the usual assumption of gaugino unification,
\beq
M_1=M_2=M_3\equiv M_{1/2}.
\label{GU}
\eeq
\item {\em Universal Higgs masses and third generation sfermions.}
  Finally, although we are not explicitly considering the mass spectrum
  in the Higgs sector and the third generation sfermions, those  mass
  parameters feed into the renormalization group equations (RGEs)
  through the hypercharge trace
\beq
S\equiv {\rm Tr}\left( YM^2 \right) \equiv
M_{H_u}^2 - M_{H_d}^2 + \sum_{i=1}^{3}
\left( M_{Q_i}^2 - 2M_{U_i}^2 + M_{D_i}^2 -M_{L_i}^2+M_{E_i}^2\right),
\label{Sdef}
\eeq
where the index $i$ now runs over the three generations of
sfermions. Extending (\ref{SU5}) over the third generation as well,
the last term in (\ref{Sdef}) is identically zero, and our last GUT
scale assumption becomes
\beq
S(M_{GUT}) = 0 \quad \Longleftrightarrow \quad M_{H_u}(M_{GUT}) =
M_{H_d}(M_{GUT}),
\label{HU}
\eeq
which is essentially the requirement of Higgs mass unification.
\end{enumerate}

\begin{table}[t]
\centering
\begin{tabular}{| c | c | c | c | c | c |}
\hline
Case      & \multicolumn{3}{c|}{Unification assumption}              & Number & Input parameters\\ \cline{2-4}
number  & Sfermion      &  Gaugino  &  Higgs  & of inputs  & at $M_{GUT}$\\
$(m)$     & (\ref{SO10})  &  (\ref{GU})  &  (\ref{HU})  & $d^{(m)}$ &   $\vec{G}^{(m)}$     \\ \hline\hline
1            &      yes           &    yes          &    yes         & 2             & $M_{0}$, $M_{1/2}$  \\
2            &       no           &    yes          &    yes         & 3             & $M_{5}$, $M_{10}$, $M_{1/2}$  \\
3            &      yes           &    yes          &    no          & 3             & $M_{0}$, $M_{1/2}$, $S$  \\
4            &       no           &    yes          &     no         & 4             & $M_{5}$, $M_{10}$, $M_{1/2}$, $S$  \\
5            &      yes           &     no          &    yes         &  4            &  $M_{0}$, $M_{1}$, $M_{2}$, $M_{3}$ \\
6            &       no           &     no          &    yes         & 5             & $M_{5}$, $M_{10}$, $M_{1}$, $M_{2}$, $M_{3}$ \\
7            &      yes           &     no          &     no         & 5             & $M_{0}$, $M_{1}$, $M_{2}$, $M_{3}$, $S$  \\
8            &       no           &     no          &     no         & 6             & $M_{5}$, $M_{10}$, $M_{1}$, $M_{2}$, $M_3$, $S$  \\
\hline
\end{tabular}
\caption{\label{tab:models}
The GUT scale assumptions behind each of the eight different SUGRA
models studied in the paper.  For completeness, we also list the
number, $d$, of GUT scale input mass parameters in each case, as well as the
names of those parameters.}
\end{table}

In this paper, we will always assume the GUT scale unification of
sfermion families in $SU(5)$ multiples described by eq.~(\ref{SU5}).
By either imposing or not imposing the remaining three GUT scale
assumptions (\ref{SO10}),
(\ref{GU}) and (\ref{HU}), we obtain a total of $2^3=8$ different model scenarios
which are listed in table~\ref{tab:models}. Each case has a different number of 
input mass parameters at the GUT scale (those parameters are listed explicitly 
in the last column of the table), so the sparticle spectrum is
constrained to varying degrees in each scenario.

We proceed to study the allowed hierarchies in each model, using the
shorthand notation from table~\ref{tab:summary}
to label each hierarchy, ordering the particles from heaviest to lightest.
For example, $GQUDLWEB$ is a hierarchy with $M_G > M_Q > M_U > M_D > M_L >
M_W > M_E > M_B$.
In all eight models from table~\ref{tab:models}, the number $d$ of input
GUT-scale parameters (listed in the fifth column) is less than the
number of measured parameters (\ref{eq:measured}).  Therefore, the
GUT-scale boundary conditions will impose certain relationships among
the low-energy parameters (\ref{eq:measured}), and as a result,
depending on the specific model, some hierarchies will be allowed,
while others will be forbidden. The main purpose of this paper is to
compile a complete catalogue of the allowed and forbidden hierarchies
in each of the eight model scenarios from table~\ref{tab:models}. Our
approach is completely general and can be easily applied to other
SUSY-breaking mechanisms, where the soft mass parameters are given by
a different set of boundary conditions at $M_{GUT}$, or are generated
at an initial scale different from $M_{GUT}$.  The method is not
limited to SUSY models, and is equally applicable to non-SUSY
scenarios, as long as the relevant mass parameters are evolved through
a linear and homogeneous system of RGEs.

This paper is one in a long line of works addressing the major
question of how to test for supersymmetry once a signal of new physics
is seen \cite{Feng:1995zd}. After the initial determination of (some
of) the SUSY mass parameters, one would like to know whether the data
is consistent with supersymmetry in general, or with a specific model
(e.g., one of the models in table~\ref{tab:models}).  Several approaches are possible:
\begin{itemize}
\item {\em Top-down approach.} One can try to fit the data directly to
  the GUT-scale input parameters of the corresponding model
  \cite{Bechtle:2012zk,Buchmueller:2012hv,Strege:2012bt}.  In doing
  so, one is faced with the usual challenges of global minimization
  problems.  In particular, one has to be careful to exhaustively
  cover all corners of parameter space, in order to be sure that a bad
  fit really excludes the model.
\item {\em Bottom-up approach.} Alternatively, one can use the
  measured SUSY parameters at the electroweak scale as boundary
  conditions and run the RGEs in reverse back to the GUT scale
  \cite{Blair:2000gy,Blair:2002pg}. This method provides a clear and
  intuitive picture of unification. However, since the RGE's are
  coupled, one needs a sufficiently large number of measurements in
  order to completely specify the initial conditions at low energies.
  Therefore, at the initial stages of the discovery, when only a
  fraction of the SUSY mass spectrum has been measured, this method
  generally does not apply. 
\item {\em SUSY mass sum rules.} The fact that the number of GUT scale
  inputs is less than the number of sparticle masses is reflected in
  the existence of certain model-dependent relations (``sum rules" for
  short) among the low energy parameters \cite{Martin:1993ft}.  This
  idea has been richly explored in many different model scenarios 
\cite{Cheng:1994bi, Yamada:1996jf, Strumia:1997xs, Matchev:1998fm,
  Huitu:2003az, Ananthanarayan:2003ca, Demir:2004aq,
  Ananthanarayan:2004dm, Kawamura:2007fx, Ananthanarayan:2007fj,
  Bhattacharya:2007dr, Bhattacharya:2008dk, Blanke:2010cm,
  Carena:2010gr, Jaeckel:2011ma, Carena:2012he, Miller:2012vn},
and has some relevance to our approach as well. For example, as
illustrated below, we often find that the reason why certain mass
hierarchies are not allowed is simply the fact that they violate one
or more of the respective SUSY mass sum rules. However, our analysis
will extend one step further and identify hierarchies which are
consistent with the sum rules, but disallowed for other reasons ---
e.g., because they would require unphysical values for the GUT-scale
boundary conditions or because they lead to tachyons in the spectrum.
\end{itemize}

The main advantages of our approach compared to these earlier studies
are the following:
\begin{enumerate}
\item We often do not require knowledge of the complete mass spectrum
  in order to decide that a given model is ruled out. In the extreme
  cases, the knowledge of just two to four mass parameters can be
  already sufficient to discredit a given model hypothesis, regardless
  of the values of the remaining mass parameters (which may even be
  unmeasured).
\item We do not require any RGE analysis post-discovery, since all the
  required analytical work has already been performed ahead of time.
\item Our results are robust and reliable in the sense that they are
  obtained analytically and do not rely on any scanning of parameter
  spaces or on numerical fitting. 
\end{enumerate}

The paper is organized as follows. In section~\ref{sec:notation} we
introduce our notations and describe the usual numerical procedure
which relates GUT-scale to weak-scale parameters.  
The next sections describe various ways of determining if 
and understanding why a hierarchy is allowed or forbidden.
In section~\ref{sec:lin-alg} we explore a linear algebraic way to
identify allowed and forbidden hierarchies.  
In section~\ref{sec:allowed} we present our results on the 
allowed sets of mass hierarchies within each theory model scenario from table~\ref{tab:models}.
Then in section~\ref{sec:forbidden} we demonstrate how the enumeration of
forbidden hierarchies is enormously simplified by the consideration 
of forbidden {\em sub-hierarchies}.
In sections~\ref{sec:msugra} and \ref{sec:CMSSMhierarchies} 
we examine the use of mass sum rules to understand intuitively the sets of 
allowed and forbidden hierarchies, employing the CMSSM as a concrete example. 
Section~\ref{sec:conclusions} is reserved for our conclusions.  
In appendix \ref{app:forbidden} we summarize our results on the sets of forbidden hierarchies.
The {\tt Python} code which contains all the hierarchy information and
which accompanies this paper is described in appendix~\ref{sec:python-code}.

\section{Notations and setup}
\label{sec:notation}

The connection between the phenomenological parameters (\ref{eq:measured})
measured at the electroweak scale $M_{EW}$
and the corresponding input parameters (specified at the GUT scale $M_{GUT}$) 
is provided by the MSSM RGEs, which allow exact analytical solutions
at one loop.  
The starting point of our analysis is therefore the map
\cite{Ramond:1999vh}
\beq
\vec{W}\equiv
\left( \begin{array}{c}
M_Q^2 \\ M_U^2 \\ M_D^2 \\ M_L^2 \\ M_E^2 \\ M_G^2 \\ M_W^2 \\ M_B^2 
\end{array}
\right)_{M_{EW}}
\equiv R^{(m)}\, \vec{G}^{(m)} |_{M_{GUT}}\, ,
\label{RGEsystem}
\eeq
where $\vec{W}$ is a vector of weak scale {\em mass-squared} parameters
and $\vec G^{(m)}$ is the corresponding vector of independent GUT scale 
{\em mass-squared} parameters for case $m$ (see the last column in table~\ref{tab:models}).
Note that the length of the vector $\vec G^{(m)}$ and the meaning of its 
components both depend on the specific SUSY scenario:
\beq
\dim\left( \vec G^{(m)} \right) = d^{(m)},
\label{dimensionofG}
\eeq
where the number $d$ of GUT scale input parameters is listed in the fifth column of table~\ref{tab:models}.
The matrix $R^{(m)}$, therefore, is an $8\times d^{(m)}$ matrix which encodes the solution to the RGEs.
It is instructive to illustrate eq.~(\ref{RGEsystem}) with a couple of examples.

First, consider the most general case of model $m=8$, when there are $d^{(8)}=6$ independent
inputs. The GUT scale parameter vector will then be\footnote{Note that for the gaugino masses 
we continue to use the notation $M_G$, $M_W$ and $M_B$.}
\beq
\vec G^{(8)} = \left(G_1^{(8)}, \, G_2^{(8)},\, G_3^{(8)},\, G_4^{(8)}, \, G_5^{(8)},\, G_6^{(8)} \right)
= \left( M_{10}^2, M_5^2, \tilde S, M_{3}^2, M_{2}^2, M_{1}^2 \right)\, ,
\eeq
where as usual we have rescaled the hypercharge trace parameter as
\beq
\tilde S =  \frac{1}{66}\left( \frac{\alpha_1}{\alpha_G}-1\right) S \, .
\eeq
In what follows, the three $\alpha_i\equiv \frac{g_i^2}{4\pi}$ parameterize the SM gauge couplings 
at the weak scale, while $\alpha_G$ stands for the corresponding unified gauge coupling at the GUT scale.
In principle, the precise values of the gauge couplings depend on the scale of sparticle masses. 
For TeV scale SUSY, however, we shall use the typical values $\alpha_G \simeq 0.041$, $\alpha_1 \simeq 0.017$, 
$\alpha_2 \simeq 0.034$, and $\alpha_3 \simeq 0.118$.  

For model $m=8$, then, Eq.~(\ref{RGEsystem}) contains the $8\times 6$ matrix $R^{(8)}$:
\beq
\vec{W}\equiv
\left( \begin{array}{c}
M_Q^2 \\ M_U^2 \\ M_D^2 \\ M_L^2 \\ M_E^2 \\ M_G^2 \\ M_W^2 \\ M_B^2 
\end{array}
\right)_{M_{EW}}
=
\left( 
\begin{array}{ccrccc}
~1~ & ~0~ &    1 & c_3 & c_2 & \frac{1}{36}c_1  \\
1 & 0 & -4 & c_3 &    0  & \frac{4}{9}c_1  \\
0 & 1 &  2 & c_3 &    0  & \frac{1}{9}c_1  \\
0 & 1 & -3 &  0    & c_2 & \frac{1}{4}c_1  \\
1 & 0 &  6 &  0    &    0  & c_1  \\
0 & 0 &    0       & a_3 &    0  &   0     \\
0 & 0 &    0       & 0  & a_2 &      0     \\
0 & 0 &    0       &  0  &   0 &  a_1        
\end{array}
\right)
\left( 
\begin{array}{c}
M_{10}^2 \\ M_5^2 \\ \tilde S \\ M_{3}^2 \\ M_{2}^2 \\ M_{1}^2 
\end{array}
\right)_{M_{GUT}} 
\equiv R^{(8)}\, \vec{G}^{(8)}\, ,
\label{Rdefinition}
\eeq
where the dimensionless coefficients $a_i$ and $c_i$ are defined as
\bea
a_i &\equiv& \frac{\alpha_i^2}{\alpha_G^2}\, ,  \label{ai} \\ [2mm]
c_i &\equiv& 
\left(
\begin{array}{r}
\frac{2}{11} \\ [1mm] \frac{3}{2} \\[1mm] -\frac{8}{9}
\end{array}
\right)
\left( 1- a_i\right).
\label{ci}
\eea
Specifically, in our numerical analysis below we shall use the values
$a_1 = 0.170$, $a_2  = 0.676$ and $a_3 = 8.29$.
Analyses performed after a SUSY discovery could use updated values and
scan over the then current experimental uncertainties on these
parameters.


As another illustration, let us also discuss a simpler case ---
model $m=3$ has three input GUT scale parameters: $M_0$, $\tilde S$ and $M_{1/2}$.
A non-zero value for $\tilde S$ at the GUT scale can be generated by non-universal 
Higgs masses, thus in what follows we shall label this model as NUHM
\cite{Berezinsky:1995cj,Nath:1997qm}. 
The GUT scale parameter vector $\vec G^{(3)}$ can be taken to be
\beq
\vec G^{(3)} = \left(G_1^{(3)}, \, G_2^{(3)},\, G_3^{(3)}\right)
= \left( M_{0}^2,\, \tilde S, \, M_{1/2}^2 \right)\, ,
\label{G3definition}
\eeq
and the corresponding $8\times 3$ matrix $R^{(3)}$ will be 
\beq
R^{(3)}=
\left( 
\begin{array}{crr}
~~1~~ & 1 & ~~c_3 + c_2 + \frac{1}{36}c_1  \\
1 & -4 & c_3  + \frac{4}{9}c_1  \\
1 &  2 & c_3  + \frac{1}{9}c_1  \\
1 &  -3 & c_2 + \frac{1}{4}c_1  \\
1 &   6 &  c_1  \\
0 &  0 & a_3      \\
0 &  0 & a_2      \\
0 &  0 & a_1        
\end{array}
\right)\, .
\label{R3definition}
\eeq


Finally, let us also mention the case of CMSSM, in which there are only two GUT scale parameters:
\beq
\vec G^{(1)} = \left(G_1^{(1)}, \, G_2^{(1)} \right)
= \left( M_{0}^2, \, M_{1/2}^2 \right)\, ,
\label{G1definition}
\eeq
and the corresponding $8\times 2$ matrix $R^{(1)}$ is given by
\beq
R^{(1)}=
\left( 
\begin{array}{cr}
~~1~~ &  c_3 + c_2 + \frac{1}{36}c_1  \\
1 &  c_3  + \frac{4}{9}c_1  \\
1 &   c_3  + \frac{1}{9}c_1  \\
1 &   c_2 + \frac{1}{4}c_1  \\
1 &    c_1  \\
0 &   a_3      \\
0 &   a_2      \\
0 &   a_1        
\end{array}
\right)\, .
\label{R1definition}
\eeq

\section{Linear algebraic approach}
\label{sec:lin-alg}

\subsection{Theory}
\label{sec:theory}

We now develop a criterion to decide whether, within a given model $m$ from table~\ref{tab:models}, 
a specific $n$-particle hierarchy is possible or not. Note that in general we consider values for $n$ in the range
\beq
2 \le n \le 8,
\label{nrange}
\eeq
i.e., we study not only full hierarchies of all 8 particles in eq.~(\ref{eq:measured}), 
but also arbitrary subsets of {\em less than} 8 particles ($n<8$). 
As an extreme example, which we shall use in this section to illustrate the formal math,
consider the 2-particle hierarchy ($n=2$) $WQ$ in the CMSSM model ($m=1$).

Recall that the weak-scale masses-squared were arranged in the vector $\vec{W}$
introduced in (\ref{RGEsystem}). Thus we can specify a given (sub-)hierarchy of interest 
by either a string of $n$ letters taken from the second row of table~\ref{tab:summary}, 
or by a sequence of $n$ integers $(i_1, i_2, \ldots , i_n)$
denoting the corresponding components of the vector $\vec{W}$
and taken from the third row of table~\ref{tab:summary}.
For example, the hierarchy $M_W>M_Q$ can be represented by the letter string
$WQ$ as explained in the introduction, or by the integer pair $(i_1,i_2)=(7,1)$
representing the respective components of the vector $\vec{W}$, namely
$(W_7,W_1)$. Thus in general an $n$-particle hierarchy $(i_1,i_2,\ldots,i_n)$
is represented by an $n$-component vector
\beq
(W_{i_1} , W_{i_2} , \cdots ,W_{i_n}).
\eeq
When we are dealing with a sub-hierarchy ($n<8$), this vector has fewer than 8 components,
and we find it convenient to always promote it back to an 8-component vector by adding the 
remaining $\vec{W}$ components $W_{j_1} , W_{j_2} , \cdots ,W_{j_{8-n}}$:
\beq
\vec  H  =\left( W_{i_1},W_{i_2},, \cdots, \, W_{i_n}, W_{j_1},
  W_{j_2}, \cdots, W_{j_{(8-n)}} \right) \, , \quad
j_k \not\in \{ i_1, \, i_2, \, \cdots, i_n\}.
\label{HeqW8}
\eeq
In other words, the first $n$ components of $\vec H$ are the weak
scale masses-squared in the $n$-particle hierarchy of interest 
(in decreasing order of mass), while the remaining $8-n$ components 
are the remaining weak scale mass parameters, taken in arbitrary order.
Operationally, we can build the vector $\vec H$ by multiplying $\vec W$ by 
the $8\times 8$ square matrix
\beq
U_{\vec{H}} =
 \begin{pmatrix}
  \delta_{(i_1,1)} &  \delta_{(i_1,2)} & \cdots &  \delta_{(i_1,8)} \\
  \delta_{(i_2,1)} &  \delta_{(i_2,2)} & \cdots &  \delta_{(i_2,8)} \\
  \vdots  & \vdots  & \ddots & \vdots  \\
  \delta_{(i_n,1)} & \delta_{(i_n,2)} & \cdots &  \delta_{(i_n,8)} \\
   \delta_{(j_1,1)} & \delta_{(j_2,2)} & \cdots &  \delta_{(j_8,8)} \\
     \vdots  & \vdots  & \ddots & \vdots  \\
    \delta_{(j_{(8-n)},1)} & \delta_{(j_{(8-n)},2)} & \cdots &  \delta_{(j_{(8-n)},8)} \\    
 \end{pmatrix} ,
\eeq
so that 
\beq
{\vec H} \equiv U_{\vec{H}}\, \vec W. 
\label{Hndefinition}
\eeq
For our example of the 2-particle hierarchy $WQ=(W_7,W_1)$, we have
$i_1=7$ and $i_2=1$. Then for concreteness we can take 
$j_1=2$, $j_2=3$, $j_3=4$, $j_4=5$, $j_5=6$ and $j_6=8$,
so that eq.~(\ref{Hndefinition}) reads
\beq
{\vec H} \equiv
\begin{pmatrix}
H_1 \\
H_2 \\
H_3 \\
H_4 \\
H_5 \\
H_6 \\
H_7 \\
H_8
\end{pmatrix}
=
\begin{pmatrix}
0 & 0 & 0 & 0 & 0 & 0 & 1 & 0 \\
1 & 0 & 0 & 0 & 0 & 0 & 0 & 0 \\
0 & 1 & 0 & 0 & 0 & 0 & 0 & 0 \\
0 & 0 & 1 & 0 & 0 & 0 & 0 & 0 \\
0 & 0 & 0 & 1 & 0 & 0 & 0 & 0 \\
0 & 0 & 0 & 0 & 1 & 0 & 0 & 0 \\
0 & 0 & 0 & 0 & 0 & 1 & 0 & 0 \\
0 & 0 & 0 & 0 & 0 & 0 & 0 & 1 \\
\end{pmatrix}
\begin{pmatrix}
W_1 \\
W_2 \\
W_3 \\
W_4 \\
W_5 \\
W_6 \\
W_7 \\
W_8
\end{pmatrix}
=
\begin{pmatrix}
W_7 \\
W_1 \\
W_2 \\
W_3 \\
W_4 \\
W_5 \\
W_6 \\
W_8
\end{pmatrix}.
\label{HnWQ}
\eeq

Viable mass spectra exhibiting the required hierarchy are those for which 
the first $n$ components of the vector $\vec H$ are all positive and in 
decreasing order, while the remaining $8-n$ components are just positive (and in any order).
Since we are interested in a hierarchical ranking of the first $n$ components, it is
convenient to define another vector, $\vec \Delta_n$, in terms of the mass-squared differences:
\bea
\vec \Delta_n \equiv
\begin{pmatrix}
H_{1}-H_{2} \\
H_{2}- H_{3} \\
\hdots \\
H_{(n-1)}-H_{n} \\
H_{n} \\
H_{n+1} \\
\hdots \\
H_{(8-n)} 
\end{pmatrix}
\equiv
D_n\,
\vec H,
\label{Deltandefinition}
\eea
where the $8\times 8$ square matrix $D_n$ is defined by
\begin{eqnarray}
D_n = 
 \begin{pmatrix}
 1 & -1 & 0 & \cdots & 0 & 0 & 0 \cdots & 0\\
 0 & 1 & -1 & \cdots & 0 & 0 & 0 \cdots & 0\\
 \vdots & \vdots & \vdots & \ddots & \vdots & \vdots & \ddots & \vdots  \\
 0 & 0 & 0 & \cdots & 1 & -1& 0 \cdots & 0 \\
 0 & 0 & 0 & \cdots & 0 & 1 & 0 \cdots & 0\\
 0 & 0 & 0 & \cdots & 0 & 0 & 1 \cdots & 0\\
  \vdots & \vdots & \vdots & \ddots & \vdots & \vdots & \ddots & \vdots  \\
 0 & 0 & 0 & \cdots & 0 &   0 & \cdots & 1
 \end{pmatrix}
=
\left\{
\begin{array}{c}
\delta_{ij}-\delta_{i+1,j}, \quad {\rm for}\ i<n;\\
\delta_{ij}, \qquad\qquad {\rm for}\ i\ge n.
\end{array}
\right.
\label{Dngendefinition}
\end{eqnarray}
For the $WQ$ example in the CMSSM, the corresponding matrix $D_2$ is
\beq
D_2 = 
 \begin{pmatrix}
 1 & -1 & 0 & 0 & 0 & 0 & 0 & 0 \\
 0 & 1 & 0 & 0 & 0 & 0 & 0 & 0  \\
 0 & 0 & 1 & 0 & 0 & 0 & 0 & 0  \\
 0 & 0 & 0 & 1 & 0 & 0 & 0 & 0  \\
 0 & 0 & 0 & 0 & 1 & 0 & 0 & 0  \\
 0 & 0 & 0 & 0 & 0 & 1 & 0 & 0  \\
 0 & 0 & 0 & 0 & 0 & 0 & 1 & 0  \\
 0 & 0 & 0 & 0 & 0 & 0 & 0 & 1  \\
 \end{pmatrix} .
\label{DnWQ}
\eeq

Combining (\ref{RGEsystem}), (\ref{Hndefinition}) and (\ref{Deltandefinition}), we get
\begin{equation}
\vec \Delta_{n,\vec{H}}^{(m)} = D_n\, U_{\vec{H}}\, R^{(m)}\, \vec{G}^{(m)} |_{M_{GUT}}.
\label{eq:simple-master}
\end{equation}
A set of GUT scale parameters $\vec G^{(m)}$ is able to produce
a physically meaningful mass spectrum for the $n$-particle
hierarchy under consideration\footnote{Note that by construction
$\vec \Delta_{n,\vec{H}}^{(m)} $ depends on the hierarchy, so that the vector $\vec \Delta_{n,\vec{H}}^{(m)} $
is defined on a case by case basis, as suggested by the index $\vec{H}$.} if and only if all eight components of
$\vec \Delta_{n,\vec{H}}^{(m)} $ are positive. We have thus reduced our original problem 
of proving the existence of the hierarchy (\ref{HeqW8})
to the problem of finding solutions such that all components of $\vec \Delta_{n,\vec{H}}^{(m)} $
are positive.

The mathematical problem, therefore, is to determine whether for the $8\times d^{(m)}$ matrix
\begin{equation}
M_{n,\vec{H}}^{(m)} = D_n\, U_{\vec{H}}\, R^{(m)}
\label{eq:A}
\end{equation}
appearing in (\ref{eq:simple-master}), 
there exists $\vec{G}^{(m)} \in {\mathbb R}^{d^{(m)}}$ such that the vector 
\beq
\vec \Delta_{n,\vec{H}}^{(m)} = M_{n,\vec{H}}^{(m)} \vec{G}^{(m)}
\label{DeltaHNm}
\eeq
has only positive components.  For this purpose, we can use 
a theorem known as Gordan's Alternative,
\cite{Gordan} 
which states that for every $r \times s$ matrix, $A$, either
\begin{equation}
\exists \, \vec x \in {\mathbb R}^s\, \mathrm{such} \,  \mathrm{that}\, A
\vec x \gg 0
\label{eq:Gordan-1}
\end{equation}
or
\begin{equation}
\exists \, \vec p \in {\mathbb R}^r \, \mathrm{such} \,
\mathrm{that}\, \vec p^\top A = 0,\, \vec p > 0.
\label{eq:Gordan-2}
\end{equation}
Here we have used the usual mathematical notation that for a vector $\vec{a}$, $\vec{a} \gg 0$ indicates that 
all components of $\vec{a}$ are positive, while $\vec{a} > 0$ only implies that all components are non-negative and in addition 
they cannot be all zero.

We can now directly make use of Gordan's alternative by identifying the matrix $A$ with our matrix
$M_{n,\vec{H}}^{(m)}$ from (\ref{eq:A}). Then, if we can find a satisfactory solution to the equation
\beq
 \vec p^\top M_{n,\vec{H}}^{(m)} = 0,
\label{pTM}
\eeq
the corresponding hierarchy is forbidden, i.e., there are no values of the 
GUT scale parameters $\vec{G}^{(m)}$ such that $\vec{\Delta}_{n,\vec{H}}^{(m)}\gg 0$.
It is often easier to determine whether a solution to
eq.~(\ref{pTM}) exists than to prove the existence of a vector 
$\vec{G}^{(m)}$ for which $M_{n,\vec{H}}^{(m)}\vec{G}^{(m)}\gg 0$.
We shall demonstrate this with a few concrete examples below in Sec.~\ref{sec:examples}.

Note that this approach also allows us to add additional constraints 
on the range of the GUT scale parameters themselves.
Since the GUT scale boundary conditions (the components of $\vec G^{(m)}$)
are masses squared, one could demand that they are positive as well.
This is certainly true for the gaugino masses squared $M_i^2$, and 
optionally for the scalar masses squared $M_{10}^2$ and $M_{5}^2$ \cite{Feng:2005ba}.
The trace $\tilde S$ can, of course, have either sign.

If we demand that a particular GUT scale parameter $G^{(m)}_i$ be positive,
we simply supplement the matrix $M_{n,\vec{H}}^{(m)}$ with a row vector $\vec{e}_i$ with components
$(\vec{e}_i)_j\equiv \delta_{ij}$.
Then the matrix $A$ becomes
\begin{eqnarray}
A = \begin{pmatrix}
M_{n,\vec{H}}^{(m)} \\
\vec{e}_i
\end{pmatrix}.
\label{eq:A-def}
\end{eqnarray}
Then,
\begin{eqnarray}
  A \vec G^{(m)} = 
  \begin{pmatrix}
    M_{n,\vec{H}}^{(m)} \vec G^{(m)} \\
    G_i^{(m)} 
  \end{pmatrix}.
  \label{eq:A-train}
\end{eqnarray}
Gordan's Alternative tells us that either one can find acceptable\footnote{Meaning $G_i^{(m)}>0$.} values 
of the GUT scale inputs which will lead to the required hierarchy,
or we can find a non-trivial solution to the matrix equation
\begin{eqnarray}
  \vec p^\top A = \vec 0,
 \label{eq:A-trans}
\end{eqnarray}
such that all components of $\vec{p}$ are non-negative.
Obviously, the procedure of enlarging the matrix $A$ as in (\ref{eq:A-def})
can be repeated for as many GUT scale parameters as necessary.

An alternative method, which may be easier to implement in practice, 
is to use the {\tt FindInstance[]} command in {\tt Mathematica} to
obtain a list of all possible $n$-particle hierarchies in each of the
eight SUSY scenarios.  These lists are provided explicitly, along with
methods to access this data, in the {\tt Python} code described in
appendix~\ref{sec:python-code}.   

\subsection{Specific examples}
\label{sec:examples}

\subsubsection{A forbidden hierarchy: $WQ$ in the CMSSM ($m=1$)}

Let us first apply the formalism of the preceding subsection to the case of the 
two-particle hierarchy $WQ$ in the CMSSM model ($m=1$).
The matrix $D_2$ was given in (\ref{DnWQ}),
the matrix $U_{\vec{H}}$ was defined in (\ref{HnWQ}), 
while $R^{(1)}$ was presented in (\ref{R1definition}).
Thus from eq.~(\ref{eq:A}) we can find the matrix $M_{2,\vec{H}}^{(1)}$:
\beq
M_{2,\vec{H}}^{(1)} = 
 \begin{pmatrix}
-1~~ &  a_2 - c_3 - c_2 - \frac{1}{36}c_1  \\
1 &  c_3 + c_2 + \frac{1}{36}c_1  \\
1 &  c_3  + \frac{4}{9}c_1  \\
1 &   c_3  + \frac{1}{9}c_1  \\
1 &   c_2 + \frac{1}{4}c_1  \\
1 &    c_1  \\
0 &   a_3      \\
0 &   a_1        
 \end{pmatrix}.
\eeq
The GUT scale parameters are given in (\ref{G1definition}). 
Requiring both to be positive, the enlarged matrix $A$ becomes
\beq
A = 
 \begin{pmatrix}
-1~~ &  a_2 - c_3 - c_2 - \frac{1}{36}c_1  \\
1 &  c_3 + c_2 + \frac{1}{36}c_1  \\
1 &  c_3  + \frac{4}{9}c_1  \\
1 &   c_3  + \frac{1}{9}c_1  \\
1 &   c_2 + \frac{1}{4}c_1  \\
1 &    c_1  \\
0 &   a_3      \\
0 &   a_1    \\
1 &  0   \\
0 &  1 \\     
 \end{pmatrix}
 =
  \begin{pmatrix}
-1~~ &  -6.29  \\
1 &  6.97  \\
1 &  6.54  \\
1 &  6.49  \\
1 &  0.52  \\
1 &  0.15  \\
0 &  8.29      \\
0 &  0.17    \\
1 &  0   \\
0 &  1 \\     
 \end{pmatrix}.
\label{AmatrixWQ} 
\eeq
Eqs.~(\ref{eq:A-trans}) now read
\beq
(p_1, p_2, \ldots , p_{10})
 \begin{pmatrix}
-1~~ &  -6.29  \\
1 &  6.97  \\
1 &  6.54  \\
1 &  6.49  \\
1 &  0.52  \\
1 &  0.15  \\
0 &  8.29      \\
0 &  0.17    \\
1 &  0   \\
0 &  1 \\     
 \end{pmatrix} =0.
\label{WQequations}
\eeq
Obviously, 
\beq
\vec{p} = 
\begin{pmatrix}
1\\
0\\
0\\
0\\
0\\
0\\
0\\
0\\
1\\
6.29
\end{pmatrix}
\eeq
is a solution to (\ref{WQequations}) and therefore, by Gordan's alternative, the hierarchy is not allowed.
This can also be seen by noting that the first row of the $A$ matrix (\ref{AmatrixWQ}) contains only negative components,
and the first equation in (\ref{DeltaHNm}) reads
\beq
\left(\Delta_{2,\vec{H}}^{(1)}\right)_1 = -\, G_1^{(1)} - 6.29\, G_2^{(1)} = - M_0^2 - 6.29 M_{1/2}^2.
\eeq
It is clear that the right-hand side cannot be positive, as long as both $M_0^2$ and $M_{1/2}^2$
are positive as well.

\subsubsection{A forbidden hierarchy: $LEU$ in the NUHM ($m=3$)}

As a more complicated example of a forbidden hierarchy, consider the $LEU$ 
sub-hierarchy in the CMSSM with non-universal Higgs masses \cite{Berezinsky:1995cj,Nath:1997qm}
(which in our notation simply means that $\tilde S$ is no longer zero).
For concreteness, we choose the remaining 5 particles to be $Q$, $D$, $G$, $W$, and $B$
and we form the extended vector $\vec{H}$ as in eq.~(\ref{HeqW8}):
\beq
\vec{H} = 
\begin{pmatrix}
M_L^2 \\
M_E^2 \\
M_U^2 \\
M_Q^2 \\
M_D^2 \\
M_G^2 \\
M_W^2 \\
M_B^2
\end{pmatrix}
.
\eeq
This defines the matrix 
\beq
U_{\vec H} = \begin{pmatrix}
0 & 0 & 0 & 1 & 0 & 0 & 0 & 0 \\
0 & 0 & 0 & 0 & 1 & 0 & 0 & 0 \\
0 & 1 & 0 & 0 & 0 & 0 & 0 & 0 \\
1 & 0 & 0 & 0 & 0 & 0 & 0 & 0 \\
0 & 0 & 1 & 0 & 0 & 0 & 0 & 0 \\
0 & 0 & 0 & 0 & 0 & 1 & 0 & 0 \\
0 & 0 & 0 & 0 & 0 & 0 & 1 & 0 \\
0 & 0 & 0 & 0 & 0 & 0 & 0 & 1 \\
\end{pmatrix}.
\label{UHLEU}
\eeq
Using (\ref{R3definition}), (\ref{Dngendefinition}) and (\ref{UHLEU}) we find 
the corresponding matrix $M_{3,\vec{H}}^{(3)}$ to be
\beq
M_{3,\vec{H}}^{(3)} = 
 \begin{pmatrix*}[r]
~~0 & ~~-9 &  c_2 - \frac{3}{4}c_1  \\
0 & 10&  -c_3  + \frac{5}{9}c_1  \\
1 & -4 & c_3  + \frac{4}{9}c_1  \\
1 &  1 &  ~~c_3  + c_2 + \frac{1}{36}c_1  \\
1 &  2 &  c_3 + \frac{1}{9}c_1  \\
0 &  0 &  a_3  \\
0 &  0 & a_2     \\
0 &  0 & a_1        
 \end{pmatrix*}
\simeq
 \begin{pmatrix*}[r]
0 & -9 & 0.37 \\
0 & 10 & -6.39 \\
1 & -4 & 6.54 \\
1 & 1 & 6.97 \\
1 & 2 & 6.49 \\
0 & 0 & 8.29 \\
0 & 0 & 0.68 \\
0 & 0 & 0.17
 \end{pmatrix*}.
\label{Mcase3}
\eeq
In this model scenario, we only demand that $M_0^2$ and $M_{1/2}^2$ be positive, while
$\tilde S$ can have either sign. Therefore, the matrix (\ref{Mcase3}) is extended with only two extra rows:
\beq
A=
 \begin{pmatrix*}[r]
0 & -9 & 0.37 \\
0 & 10 & -6.39 \\
1 & -4 & 6.54 \\
1 & 1 & 6.97 \\
1 & 2 & 6.49 \\
0 & 0 & 8.29 \\
0 & 0 & 0.68 \\
0 & 0 & 0.17 \\
1 & 0 & 0 \\
0 & 0 & 1
 \end{pmatrix*}
\eeq
and the system of equations (\ref{eq:A-trans}) to be studied is
\beq
(p_1, p_2, \ldots , p_{10})
 \begin{pmatrix*}[r]
0 & -9 & 0.37 \\
0 & 10 & -6.39 \\
1 & -4 & 6.54 \\
1 & 1 & 6.97 \\
1 & 2 & 6.49 \\
0 & 0 & 8.29 \\
0 & 0 & 0.68 \\
0 & 0 & 0.17 \\
1 & 0 & 0 \\
0 & 0 & 1
 \end{pmatrix*} = 0.
\eeq
This system admits strictly positive non-trivial solutions, for example
\beq
p_2 = 0.9\, p_1, \qquad p_{10} = \left( 0.9 c_3 - c_2 + 0.25 c_1\right) p_1 \simeq 5.38\, p_1 , \label{psolutionm3}
\eeq
for any positive value of $p_1$. According to Gordon's alternative,
the existence of the solution (\ref{psolutionm3}) implies that the hierarchy 
$LEU$ is forbidden in the $m=3$ model. We shall see below that $LEU$ 
is indeed one of the 36 forbidden hierarchies in this model, see eq.~(\ref{m3n3}).

\subsubsection{An allowed hierarchy: $GBL$ in the NUHM ($m=3$)}

We conclude this section with an example of an allowed hierarchy.
Consider the sub-hierarchy $GBL$, which is one of the $198$ allowed 
$3$-particle hierarchies in model $m=3$ (cf. table~\ref{tab:allowed}).  
As in the previous subsection, in model $m=3$ the input parameters are still
those given in (\ref{G3definition}), and the $R^{(3)}$ matrix is given by (\ref{R3definition}).
For concreteness, we choose the remaining 5 particles to be $Q$, $U$, $D$, $E$, and $W$
and form the extended vector $\vec{H}$ as in eq.~(\ref{HeqW8}):
\beq
\vec{H} = 
\begin{pmatrix}
M_G^2 \\
M_B^2 \\
M_L^2 \\
M_Q^2 \\
M_U^2 \\
M_D^2 \\
M_E^2 \\
M_W^2
\end{pmatrix},
\eeq
which defines the matrix
\beq
U_{\vec H} = \begin{pmatrix}
0 & 0 & 0 & 0 & 0 & 1 & 0 & 0 \\
0 & 0 & 0 & 0 & 0 & 0 & 0 & 1 \\
0 & 0 & 0 & 1 & 0 & 0 & 0 & 0 \\
1 & 0 & 0 & 0 & 0 & 0 & 0 & 0 \\
0 & 1 & 0 & 0 & 0 & 0 & 0 & 0 \\
0 & 0 & 1 & 0 & 0 & 0 & 0 & 0 \\
0 & 0 & 0 & 0 & 1 & 0 & 0 & 0 \\
0 & 0 & 0 & 0 & 0 & 0 & 1 & 0 \\
\end{pmatrix}.
\label{UHGBL}
\eeq
Then, using (\ref{R3definition}), (\ref{Dngendefinition}), (\ref{UHGBL}) and (\ref{eq:A}),
we obtain the enlarged matrix (\ref{eq:A-def})
\beq
A = \begin{pmatrix*}[r]
 0 & 0 & a_3 - a_1 \\
 -1 & 3 & ~~a_1 -c_2- \frac{1}{4}c_1 \\
 1 & -3 & c_2 + \frac{1}{4}c_1\\
 1 & 1 & c_3+c_2+\frac{1}{36}c_1 \\
 1 & -4 & c_3+\frac{4}{9}c_1 \\
 1 & 2 & c_3+\frac{1}{9}c_1 \\
 1 & 6 & c_1 \\
 0 & 0 & a_2 \\
 1 & 0 & 0 \\
 0 & 0 & 1
 \end{pmatrix*}
 \simeq
\begin{pmatrix*}[r]
 0 & 0 & 8.12 \\
 -1 & 3 & -0.35 \\
 1 & -3 & 0.52\\
 1 & 1 & 6.97 \\
 1 & -4 & 6.54 \\
 1 & 2 & 6.49 \\
 1 & 6 & 0.15 \\
 0 & 0 & 0.68 \\
 1 & 0 & 0 \\
 0 & 0 & 1
 \end{pmatrix*}.
\eeq
The corresponding linear system of equations (\ref{eq:A-trans}) is
\beq
\left(
\begin{array}{rrrrrrrrrr}
 0 & -1 & 1  &  1 & 1 & 1 & 1  & 0 & 1   & 0\\
 0 & 3  & -3 & 1 & -4 & 2 & 6 & 0 & 0   & 0 \\
8.12 & ~-0.35 & ~0.52 & ~6.97 & ~6.54 & ~6.49 & ~0.15 & ~0.68 & ~~0  & ~1
 \end{array}
 \right)
 \begin{pmatrix}
 p_1 \\
 p_2 \\
 \vdots \\
 p_{10}
 \end{pmatrix} =0 \, .
\label{eq:alt-1}
\eeq
We want to show that this system has no solutions with $\vec{p} > 0$.
To easily see this, multiply the last equation by 4 and add to the first two equations,
obtaining the equation
\bea
&& 32.47\, p_1+ 0.58\, p_2 + 0.09\, p_3 + 29.87\, p_4+23.18\, p_5 \nonumber \\
&+& 28.97\, p_6+ 7.60\, p_7+ 2.70\, p_8
+ p_9+ 4\, p_{10}=0. 
\label{eq:app-eq}
\eea
As all of the coefficients in eq.~(\ref{eq:app-eq}) are positive, 
there is clearly no valid solution for $\vec{p}$, and thus the hierarchy 
is allowed.

\section{Results: allowed $n$-particle (sub-)hierarchies}
\label{sec:allowed}

For each of the eight model scenarios in table~\ref{tab:models}, only certain sets of 
hierarchies are allowed, in the sense that there exist values for the GUT scale parameters $\vec{G}^{(m)}$
which will give rise to the given mass ordering at the weak scale. Note that in searching for viable 
hierarchies, we are interested not only in obtaining the masses in the specified ordering,
but in addition we require that all masses are physical, i.e., that all mass squared parameters 
at the weak scale (\ref{RGEsystem}) are positive. This is why the hierarchy vector $\vec{H}$
was defined in eq.~(\ref{HeqW8}) in terms of all eight components of the vector $\vec{W}$ from (\ref{RGEsystem}).

As indicated in (\ref{nrange}), we shall also be interested in $n$-particle {\em sub-hierarchies},
i.e., sets of just a few particles with $n<8$. In doing so, we are motivated by the experimental 
reality --- during the phase of initial discovery of supersymmetry, it is very likely that only a 
few superpartners will be seen. Therefore, at that stage it would make no sense to ask questions 
involving the unseen yet superpartners. Instead, one should focus on the question, 
given what has been observed so far, is the data consistent with a given  theory model or not?

\begin{table}
\centering
\begin{tabular}{{|c||r|r|r|r|r|r|r|}} \hline
 Case    & \multicolumn{7}{c|}{$n$}  \\
               \cline{2-8}
$(m)$         & 2  & 3  & 4  & 5  & 6  & 7 & 8 \\ \hline \hline 
1 & 36  & 104  & 190      & 216  &     148  &       56 & 9 \\ \hline
2 & 42  & 154  & 382      & 604  &     570  &     290 & 61 \\ \hline
3 & 49  & 198  & 519      & 852  &     827  &     430 & 92 \\ \hline
4 & 51  & 240  & 757    & 1536  &   1887 &   1252  & 340  \\ \hline
5 & 55  & 312  & 1277  & 3232  &   4560  &   3254 & 913 \\ \hline
6 & 56  & 330  & 1521  & 4806  &   8684 &   7824  & 2699 \\ \hline
7 & 56  & 336  & 1591  & 5234  & 10240  & 10224 & 3940 \\ \hline
8 & 56  & 336  & 1648  & 6028  & 13778 & 16502  & 7766 \\ \hline\hline
$8!/(8-n)!$ & 56   &  336  & 1680  & 6720  & 20160  & 40320  & 40320 \\ \hline
\end{tabular}
\caption{\label{tab:allowed}  The number of allowed $n$-particle
  hierarchies for each of the eight models considered in
  table~\ref{tab:models}.
}
\end{table}

We shall therefore allow $n$ to vary within the full range (\ref{nrange}). In general, 
for a given $n$, the number of all possible hierarchies which can be formed out of the set of
eight observables (\ref{eq:measured}) is given by $\frac{8!}{(8-n)!}$ and listed in the last row of
table~\ref{tab:allowed}. The remaining rows of the table give the number of {\em allowed}
$n$-particle hierarchies (in the sense described above), for each of the eight theoretical 
model scenarios from table~\ref{tab:models}.

Table~\ref{tab:models} reveals a significant reduction in the number of possible hierarchies. 
Consider, for example, the least constrained model, Case 8, where there are as many 
as 6 input parameters at the GUT scale. Even then, out of the $8!=40,380$ 8-particle permutations,
only $7,766$ hierarchies are possible (a little over $19\%$). The other, more constrained models, 
exhibit a further reduction of the possible mass hierarchies. In the extreme case of the CMSSM 
(model $m=1$ in table~\ref{tab:models}), there are only nine 8-particle possibilities, as explained
in Section~\ref{sec:msugra} below.

When we consider sub-hierarchies with $n<8$, the reduction in the number of allowed
sub-hierarchies in table~\ref{tab:allowed} is not as dramatic. This is because, when asking the question 
``Is this sub-hierarchy of $n$ particles allowed or not?" we allow the remaining (unseen) 
$8-n$ particles to have arbitrary masses. In particular, in models $m=7$ and $m=8$, all possible 
2-particle and $3$-particle hierarchies are still represented in the table, in spite of the 
significant reduction in terms of the number of allowed 8-particle hierarchies.

Of course, table~\ref{tab:models} only tallies up the number of possible hierarchies in each model, 
but does not reveal {\em which} particular hierarchies are allowed. Given the large numbers seen in
the table, giving here the complete list of all allowed hierarchies is impractical --- the interested reader
can easily generate the sets of allowed hierarchies from the accompanying code described in
appendix~\ref{sec:python-code}. This code can also answer queries about individual hierarchies 
(as well as sub-hierarchies) of interest.

Nevertheless, one may still wonder whether it is possible to somehow display the complete 
information about the sets of allowed hierarchies which are hiding behind table~\ref{tab:allowed}. 
In the next section~\ref{sec:forbidden}, we shall show that there is an elegant way of encoding 
and presenting the same amount of information. The key idea is, instead of studying the number of
allowed hierarchies (which is typically very large, as seen in table~\ref{tab:allowed}), to 
focus on the (much smaller) number of forbidden hierarchies, 
{\em starting with the smallest possible values of $n$}.

\subsection{The set of allowed hierarchies in the CMSSM}
\label{sec:msugra}

Table~\ref{tab:models} revealed that in each of the eight model 
scenarios considered, the number of GUT scale input parameters 
$d^{(m)}$ is less than the number (eight) of weak scale parameters 
under consideration. Hence the hierarchies must satisfy $8-d^{(m)}$ sum rules, 
which explains the reduction in the number of allowed hierarchies observed in table~\ref{tab:allowed}.

We shall now illustrate how these sum rules help to identify allowed hierarchies 
using the example of the CMSSM ($m=1$). In this case there are  two input parameters
($M_0^2$ and $M_{1/2}^2$), so we have the following six sum rules:
\bea
%
&& (5 c_1-12 c_2) (M_D^2-M_U^2) -4 c_1 (M_Q^2-M_U^2) = 0 \, , \label{sugrasr1} \\ [2mm]
&& 3(5 c_1-12 c_2)(M_L^2-M_U^2) - (7 c_1 -36 c_2+36 c_3) (M_Q^2-M_U^2)  =0 \, ,\label{sugrasr2} \\ [2mm]
&& 3 (5 c_1-12 c_2)(M_E^2-M_U^2) + (20 c_1-36 c_3)(M_Q^2-M_U^2) =0 \, , \label{sugrasr3} \\ [2mm]
&& (5 c_1-12 c_2)M_G^2 + 12 a_3 (M_Q^2-M_U^2)  =0 \,, \label{sugrasr4} \\ [2mm]
&& (5 c_1-12 c_2) M_W^2 + 12 a_2 (M_Q^2-M_U^2)  = 0\, , \label{sugrasr5} \\ [2mm]
&& (5 c_1 -12 c_2) M_B^2+ 12 a_1 (M_Q^2-M_U^2) = 0. \label{sugrasr6}
\label{eq:sumR}
\eea
In obtaining these sum rules, we first used the first two rows of $R^{(1)}$ in (\ref{R1definition})
to express the GUT scale parameters, $M_0^2$ and $M_{1/2}^2$, 
in terms of the weak scale parameters, $M_Q^2$ and $M_U^2$, 
\bea
M_0^2 &=& \frac{4 ( 4c_1+9 c_3)}{15 c_1-36 c_2} M_Q^2 
- \frac{c_1+ 36 c_2 + 36 c_3}{15 c_1-36 c_2} M_U^2\, , \label{M0reexpress} \\ [2mm]
  M_{1/2}^2 &=& \frac{12}{5 c_1-12 c_2} \left(M_U^2- M_Q^2\right) .
\label{M12reexpress}
\eea
Then, the remaining six rows of the matrix $R^{(1)}$ result in the sum rules (\ref{sugrasr1}-\ref{sugrasr6}).

\begin{figure}[t]
\begin{center}
\includegraphics[width=0.7\textwidth]{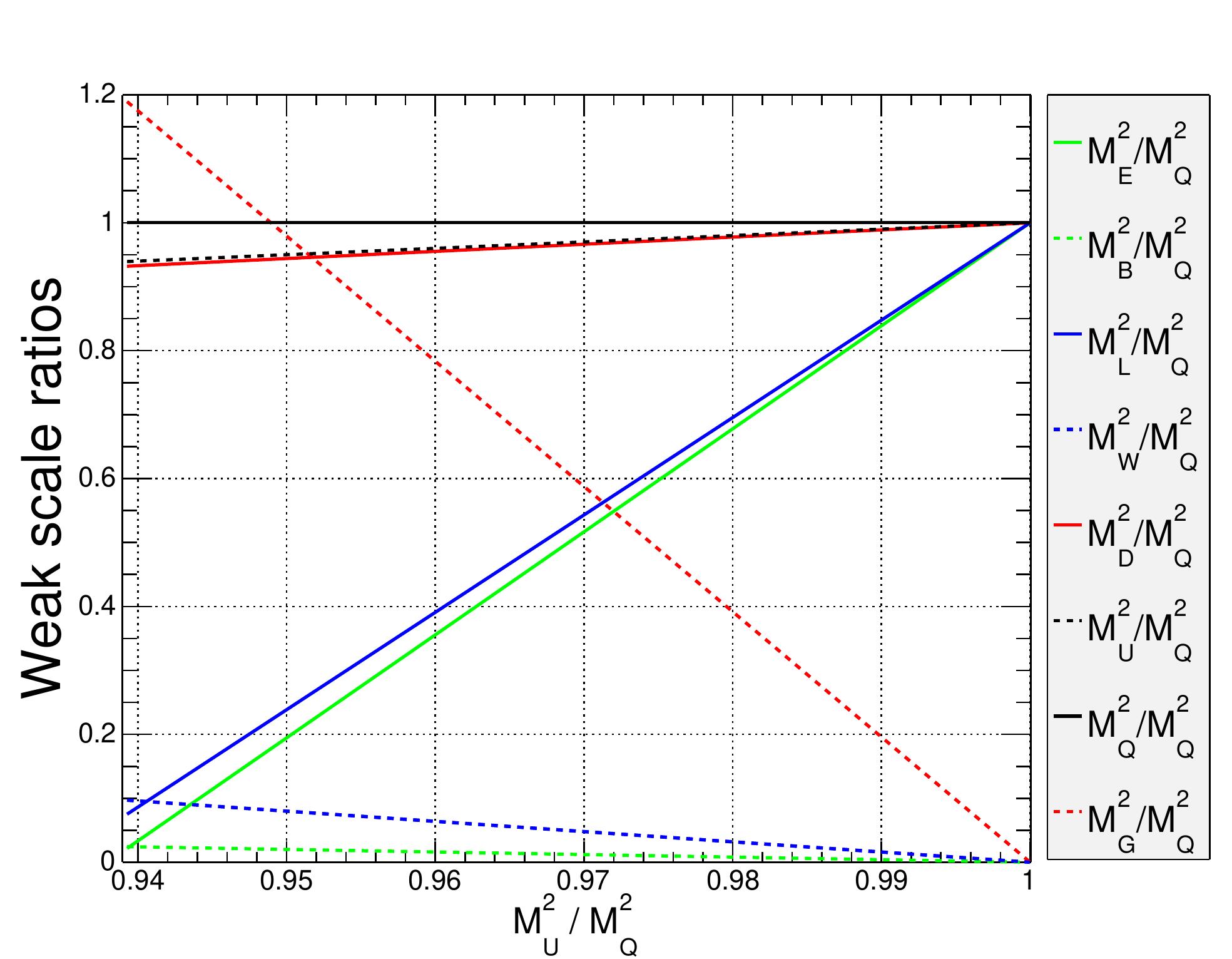} 
\end{center}
\vspace{-20 pt}
\caption{Dependence of the squared mass ratios
(\ref{eq:sumR-1}-\ref{eq:sumR-6}) on the parameter $\frac{M_{U}^2}{M_Q^2}$.
We only show the physically viable range.
\label{fig:case2}}
\end{figure}

We note that the overall mass scale does not have any impact on 
whether a given hierarchy is allowed or not.  We can therefore remove one degree
of freedom; here we choose this degree of freedom to be $M_Q^2$.   
Since the CMSSM has only two input parameters (i.e., $d^{(1)}=2$),
removing the overall scale as $M_Q^2$ leaves us with only one relevant
degree of freedom, which we can take to be the ratio $\frac{M_{U}^2}{M_Q^2}$.
The sum rules (\ref{sugrasr1}-\ref{sugrasr6}) can then be rewritten as
\bea
\frac{M_D^2}{M_Q^2}
&=&
 \frac{c_1-12 c_2} {5 c_1-12 c_2} 
\bigg(\frac{M_{U}^2}{M_Q^2}\bigg)  
+ \frac{4 c_1} {5 c_1-12 c_2}\, , \label{eq:sumR-1} \\
\frac{M_L^2}{M_Q^2} 
&=&
\frac{8 c_1 - 36 c_3}{15 c_1-36
  c_2}
\bigg(\frac{M_{U}^2}{M_Q^2}\bigg)
+ \frac{7 c_1 -36 c_2+36 c_3}{15 c_1-36 c_2} \, ,\label{eq:sumR-2} \\
\frac{M_E^2}{M_Q^2}
&=&
\frac{35 c_1 - 36 c_2 - 36 c_3}{15 c_1 - 36 c_2} 
\bigg(\frac{M_{U}^2}{M_Q^2}\bigg) 
- \frac{20 c_1 - 36 c_3}{15 c_1 - 36 c_2} \, , \label{eq:sumR-3} \\ 
\frac{M_G^2}{M_Q^2} 
&=&
\frac{12 a_3}{5 c_1-12 c_2}
\bigg(\frac{M_{U}^2}{M_Q^2}\bigg) 
-\frac{12 a_3}{5 c_1-12 c_2} \, ,\label{eq:sumR-4} \\
\frac{M_W^2}{M_Q^2} 
&=& 
\frac{12 a_2}{5 c_1-12 c_2}
\bigg(\frac{M_{U}^2}{M_Q^2}\bigg) 
-\frac{12 a_2}{5 c_1-12 c_2}\, , \label{eq:sumR-5} \\
\frac{M_B^2}{M_Q^2} 
&=& 
\frac{12 a_1}{5 c_1 -12 c_2}  
\bigg(\frac{M_{U}^2}{M_Q^2}\bigg)
-\frac{12 a_1}{5 c_1 -12 c_2}\, , 
\label{eq:sumR-6}
\eea
which are clearly linear equations in $\frac{M_{U}^2}{M_Q^2}$.  
In general, the ratio $\frac{M_{U}^2}{M_Q^2}$ can take values in the interval $(0,\infty)$.
However, not all values will lead to physically viable mass spectra, as  
illustrated in Figure~\ref{fig:case2}. 
In particular, some of the weak scale masses squared can become negative and lead to tachyonic particles.
As observed in in Figure~\ref{fig:case2}, the most stringent such restriction arises from the constraint 
$M_E^2>0$. Using eq.~(\ref{eq:sumR-3}) this translates into 
\beq
\frac{M_{U}^2}{M_Q^2} > \frac{20c_1-36 c_3}{35c_1-36 c_2-36c_3} \approx 0.94   \, .
\label{eq:rangeleft}
\eeq
In addition, we should also require positivity of the gaugino mass squared boundary condition
(\ref{M12reexpress}), which in turn implies
\beq
\frac{M_{U}^2}{M_Q^2} <1.
\label{rangeright} 
\eeq
Requiring that $M_0^2>0$ does not lead to any further restrictions, so that 
the physically allowed range for $\frac{M_{U}^2}{M_Q^2}$,
depicted in Figure~\ref{fig:case2}, is $(0.94, 1)$.

For any given value of $\frac{M_{U}^2}{M_Q^2}$, the
hierarchy of weak scale masses is uniquely determined.  
Furthermore, whenever a pair of lines in figure~\ref{fig:case2} cross, 
the hierarchy changes, since the mass ordering of the corresponding particles is reversed.
Therefore, we can enumerate the hierarchies by examining the 
values of the independent degree of freedom $\frac{M_{U}^2}{M_Q^2}$
at which lines in Figure~\ref{fig:case2} cross.  
Let 
\beq
\left. \frac{M_{U}^2}{M_Q^2}\right|_{M_X=M_Y}
\label{defcross}
\eeq
denote the value of $\frac{M_U^2}{M_Q^2}$ at which the 
$\frac{M_X^2}{M_Q^2}$ line and the $\frac{M_Y^2}{M_Q^2}$ 
line in Figure~\ref{fig:case2} intersect. It is easy to see that there 
are a total of eight intersection points, ordered as
\bea
&&
\left. \frac{M_{U}^2}{M_Q^2}\right|_{M_B=M_E} <
\left. \frac{M_{U}^2}{M_Q^2}\right|_{M_W=M_L} <
\left. \frac{M_{U}^2}{M_Q^2}\right|_{M_W=M_E} <
\left. \frac{M_{U}^2}{M_Q^2}\right|_{M_G=M_Q} < \nonumber \\
&<&
\left. \frac{M_{U}^2}{M_Q^2}\right|_{M_G=M_U} <
\left. \frac{M_{U}^2}{M_Q^2}\right|_{M_G=M_D} <
\left. \frac{M_{U}^2}{M_Q^2}\right|_{M_G=M_L} <
\left. \frac{M_{U}^2}{M_Q^2}\right|_{M_G=M_E}  \, .
\label{intersectionpoints}
\eea
The eight points (\ref{intersectionpoints}) divide the allowed
$\frac{M_{U}^2}{M_Q^2}$ interval into 9 sub-intervals, 
where each sub-interval corresponds to one specific hierarchy.
This confirms that the CMSSM indeed has 9 possible 8-particle hierarchies,
as shown in Table~\ref{tab:allowed}.
Figure~\ref{fig:case2} reveals that they are (going from left to right):
\bea
&GQUDWLBE:&\quad M_G > M_Q > M_U > M_D > M_W > M_L > M_B > M_E, \label{cmssmallowed-1} \\
&GQUDWLEB:&\quad M_G > M_Q > M_U > M_D > M_W > M_L > M_E > M_B, \label{cmssmallowed-2}  \\
&GQUDLWEB:&\quad M_G > M_Q > M_U > M_D > M_L > M_W > M_E > M_B, \label{cmssmallowed-3}  \\
&GQUDLEWB:&\quad M_G > M_Q > M_U > M_D > M_L > M_E > M_W > M_B, \label{cmssmallowed-4}  \\
&QGUDLEWB:&\quad M_Q > M_G > M_U > M_D > M_L > M_E > M_W > M_B, \label{cmssmallowed-5}  \\
&QUGDLEWB:&\quad M_Q > M_U > M_G > M_D > M_L > M_E > M_W > M_B, \label{cmssmallowed-6}  \\
&QUDGLEWB:&\quad M_Q > M_U > M_D > M_G > M_L > M_E > M_W > M_B, \label{cmssmallowed-7}  \\
&QUDLGEWB:&\quad M_Q > M_U > M_D > M_L > M_G > M_E > M_W > M_B, \label{cmssmallowed-8}  \\
&QUDLEGWB:&\quad M_Q > M_U > M_D > M_L > M_E > M_G > M_W > M_B. \label{cmssmallowed-9} 
\eea

\section{Results: the set of forbidden $n$-particle (sub-)hierarchies}
\label{sec:forbidden}

Table~\ref{tab:allowed} may seem an unedifying m\'{e}lange of data.
However, it turns out that if 
\begin{enumerate} 
\item one considers forbidden hierarchies rather than allowed hierarchies, and 
\item inspects a forbidden $n$-particle hierarchy for forbidden $2$, $3$, ..., $n-1$-particle sub-hierarchies contained in it, 
\end{enumerate}
then the picture becomes much simpler.  Our results in terms of {\em forbidden}
hierarchies are shown in table~\ref{tab:forbidden}.
\begin{table}
\centering
\begin{tabular}{|c||r|r|r|r|r|r|r|} \hline 
 Case  &   \multicolumn{7}{c|}{$n$} \\ \cline{2-8}
($m$)  &   2  & 3 & 4  & 5  & 6  & 7 & 8 \\ \hline \hline 
1 & 20  &   0  &   12  &       0  &       0   &        0  &       0    \\ \hline
2 & 14  &   0  &   13  &       0  &       2  &         0  &       0    \\ \hline
3 &   7  & 36  &   34  &     10  &       3  &         0  &       0    \\ \hline
4 &   5  & 18  &   79  &     14  &       3  &         0  &       0    \\ \hline
5 &   1  &   6  & 121  &   168  &   149  &       12  &       0    \\ \hline
6 &   0  &   6  &   45  &   216  &   288  &       98  &       3    \\ \hline
7 &   0  &   0  &   89  &   176  &   426  &     434  &     22    \\ \hline
8 &   0  &   0  &   32  &   148  &   809  &     398  &     54    \\ \hline
\end{tabular}
\caption{\label{tab:forbidden}  The number 
of irreducible forbidden $n$-particle hierarchies for 
each of the eight models considered in table~\ref{tab:models}.
}
\end{table}
The table is constructed as follows. For each model, we begin by considering the 
$8\times 7=56$ possible 2-particle hierarchies.  For each of these 2-particle sub-hierarchies, 
we check whether it is present in at least one of the allowed 8-particle hierarchies 
counted in the last column of table~{\ref{tab:allowed}}.\footnote{Equivalently, we can directly 
check whether the 2-particle sub-hierarchy under consideration is one of the allowed 2-particle 
sub-hierarchies from the $n=2$ column of table~{\ref{tab:allowed}}.} If the answer is ``yes", then 
the 2-particle sub-hierarchy is in principle allowed in this model, but if the answer is ``no",
this 2-particle sub-hierarchy will never appear in the model, and is therefore 
categorized as ``forbidden". The total number of forbidden 2-particle sub-hierarchies is 
then tallied up and listed in the $n=2$ column of table \ref{tab:forbidden},
while the explicit lists of the forbidden 2-particle hierarchies appear in Appendix~\ref{app:forbidden},
see eqs.~(\ref{m1n2}), (\ref{m2n2}), (\ref{m3n2}), (\ref{m4n2}), and (\ref{m5n2}).
Clearly, for any given model $m$, 
a specific 2-particle sub-hierarchy is either allowed or forbidden, so 
the total number $N^{(m)}_{2,forbidden}$ of forbidden $n=2$ sub-hierarchies in the model and 
the total number $N^{(m)}_{2,allowed}$ of allowed $n=2$ sub-hierarchies in the model 
add up to the total number of 2-particle sub-hierarchies:
\beq
N^{(m)}_{2,forbidden} + N^{(m)}_{2,allowed} = 56,
\eeq
which is seen to hold for the results in the respective columns in tables~\ref{tab:allowed} and \ref{tab:forbidden}.
We also observe that as the number $d^{(m)}$ of GUT-scale inputs increases, the number 
$N^{(m)}_{2,forbidden}$ of forbidden 2-particle sub-hierarchies decreases. In particular, for models 
6, 7, and 8,  we find that all 2-particle sub-hierarchies are possible. At the same time, in the most constrained model,
the CMSSM, there are 20 forbidden 2-particle hierarchies, which will be discussed in detail in Sec.~\ref{sec:CMSSMhierarchies}.

Having thus determined the set of all forbidden 2-particle sub-hierarchies, we now move on
to finding the set of all forbidden 3-particle sub-hierarchies. Clearly, a large number of 
3-particle sub-hierarchies will be forbidden simply because they contain a forbidden 2-particle sub-hierarchy.
Such cases are uninteresting to us, and we focus only on the search for {\em irreducible} 
forbidden 3-particle hierarchies, i.e., 3-particle sub-hierarchies which are composed of 
allowed 2-particle sub-hierarchies only. The number $N^{(m)}_{3,forbidden}$ 
of such ``newly forbidden" 3-particle sub-hierarchies is listed in the third column 
of table~\ref{tab:forbidden}.  Interestingly, the results for $N^{(m)}_{3,forbidden}$
do not follow quite the same pattern that we already observed for $N^{(m)}_{2,forbidden}$.
In particular, our 8 models fall into one of the following three categories:
\begin{itemize}
\item Models in which in principle there exist forbidden 3-particle sub-hierarchies, but 
none of them is ``newly forbidden". In other words, all forbidden 3-particle sub-hierarchies
already contain a forbidden 2-particle sub-hierarchy. According to table~\ref{tab:forbidden},
models $m=1$ and $m=2$ are of this type.
\item Models in which there are no forbidden 3-particle sub-hierarchies to begin with.
These are the models with a relatively large number of input parameters, namely models
$m=7$ and $m=8$.
\item Lastly, there are models in which we obtain non-trivial results at the level of 3-particle hierarchies.
In models $m=3$, $m=4$, $m=5$ and $m=6$, there arise newly forbidden 3-particle sub-hierarchies,
which cannot be explained simply with the presence of a forbidden 2-particle sub-hierarchy.
The number of ``newly forbidden" 3-particle sub-hierarchies does follow the expected pattern ---
generally speaking, there are more such hierarchies for the models with fewer input parameters.
\end{itemize}

Having accounted for the forbidden 2-particle sub-hierarchies and the newly forbidden 
3-particle sub-hierarchies, we then proceed to investigate the number $N^{(m)}_{4,forbidden}$
of newly forbidden 4-particle sub-hierarchies. The results are listed in the fourth column of 
table~\ref{tab:forbidden}. Unlike the case of 3-particle sub-hierarchies, we find that now all models
predict newly forbidden (4-particle) sub-hierarchies. In the case of models $m=7$ and $m=8$, this is 
the first time we encounter forbidden sub-hierarchies.

The process just described continues until we fill out the whole table~\ref{tab:forbidden}.
Comparing our results in table~\ref{tab:allowed} for the {\em allowed} sub-hierarchies
and the complementary results in table~\ref{tab:forbidden} for the {\em forbidden}
sub-hierarchies, we see significant simplification in the latter case in terms of counting
hierarchies and bookkeeping. At the same time, it is worth emphasizing that
tables~\ref{tab:allowed} and \ref{tab:forbidden} are simply two different ways to present
our results on the restrictions imposed by various models on particle mass hierarchies.
Whether we use the language of allowed or forbidden hierarchies, the amount of 
presented information remains the same.

However, in practice there may be situations where one language is preferred over the other.
For example, consider an initial discovery in which {\em only a few} supersymmetric particles 
have been observed. Let us focus on the simplest and most likely scenario where 
only two particles are detected: a lighter particle $X$ with mass $M_X$, and a heavier partner 
$Y$ with mass $M_Y$ ($M_Y>M_X$). Given just this single piece of information, what can one 
conclude about the various models from table~\ref{tab:models}? The answer is given in 
table~\ref{forbiddenmodels}, which lists all possible 2-particle hierarchies, and the corresponding 
models in which a particular hierarchy is forbidden. There are several conclusions which can be drawn from the table:
\begin{table}
\centering
\begin{tabular}{|c|c||c|c|c|c|c|c|c|c|} \hline
\multicolumn{2}{|c||}{$YX$}  & \multicolumn{8}{c|}{$X$}  \\ \cline{3-10}
\multicolumn{2}{|c||}{$(M_Y>M_X)$}  
            &  $Q$    & $~~U~~$    & $~D~$  & $~~L~~$  & $~~E~~$  & $G$ & $W$ &  $~B~$  \\ \hline \hline 
& $Q$ & - 		&           &         &          &          &         &         &           \\ \cline{2-10}
& $U$ & 1,2		&   -   &          &          &        &           &         &    \\ \cline{2-10}
& $D$ &  1     		& 1,5     &   -  &          &         &           &          &  \\ \cline{2-10}
$Y$ & 
    $L$ & 1   		& 1	    &  1,2   &   -  &         &           &          & \\ \cline{2-10}
& $E$ & 1,2   		&  1,2    &   1    &     1    &  -   &            &          &  \\ \cline{2-10}
& $G$ &      		&   	    &         &          &           &  -      &           &  \\ \cline{2-10}
& $W$ & 1,2,3,4 & 1,2     & 1,2,3 &          &           & 1,2,3,4     &  -     & \\ \cline{2-10}
& $B$ &  1,2,3,4 & 1,2     &  1,2,3 &  1,2     &           & 1,2,3,4    & 1,2,3,4  &   -\\ \hline
\end{tabular}
\caption{\label{forbiddenmodels}  The 56 possible 2-particle hierarchies $YX$ (with $M_Y>M_X$), where
the heavier (lighter) particle is shown in the corresponding row (column) of the table.
Each box in the table (except for the boxes along the diagonal) represents one of the 56 possible 
2-particle hierarchies, and the 
entries inside each box indicate the model numbers $m$ in which that hierarchy is forbidden.
}
\end{table}
\begin{itemize}
\item Models $m=6$, $m=7$ and $m=8$ are completely absent from table~\ref{forbiddenmodels}, 
which simply reflects the fact that in table~\ref{tab:forbidden} we did not find any forbidden 2-particle 
hierarchies for those models. This means that the discovery of just two supersymmetric particles
(no matter what they are) is insufficient to rule out SUSY models with large number of parameters
like models $m=6,7,8$. 
\item There are 36 2-particle sub-hierarchies which are present in all models --- the 28 cases 
above the diagonal in table~\ref{forbiddenmodels}, plus 8 more cases below the diagonal.
If the data happens to be one of those 2-particle hierarchies, we will not be able to make any definitive statements
about the eight theory models since they will all still be allowed.
\item The 2-particle hierarchies in which the heavier particle is colored while the lighter particle is not,
are not restrictive at all. This is a well-known feature of SUGRA-type scenarios, in which the 
RGE evolution tends to split the spectrum in such a way that the colored superpartners are heavier than their
electroweak counterparts. In particular, in all eight models, the gluino can be the heavier of the two particles.
Similarly, in all models, the right-handed selectron can be the lighter of the two particles.
\item Among the most restrictive 2-particle hierarchies are those in which the lighter particle is colored, while the 
heavier particle is not. For example, observing $WQ$, $BQ$, $WG$ or $BG$ rules out four of the models right away.
If both particles happen to be colored (or both are uncolored), then restrictive 2-particle hierarchies 
are those in which the lighter particle is charged under $SU(2)$ while the heavier one is not, 
e.g., $BW$, $BL$, $UQ$, $DQ$.
\end{itemize}

The python code described in appendix~\ref{sec:python-code} not only generates all
allowed hierarchies described in the previous section, but it can also easily reproduce
the results on forbidden hierarchies discussed in the current section.
For details on how to use the code and its functionality we refer the reader to 
appendix~\ref{sec:python-code}. We shall conclude this section by considering a few 
illustrative examples which allow analytical treatment by means of sum rules.

\subsection{Identifying all forbidden hierarchies in the CMSSM}
\label{sec:CMSSMhierarchies}

As seen in table~\ref{tab:forbidden}, the CMSSM is characterized by 20 forbidden 
2-particle sub-hierarchies and 12 newly forbidden 4-particle sub-hierarchies.
These forbidden sub-hierarchies are explicitly listed in eqs.~(\ref{m1n2})
and (\ref{m1n4}), respectively. Here we shall demonstrate how to re-derive 
those results with the help of the method of mass sum rules \cite{Martin:1993ft}.

Our starting point is the relation between the weak-scale and GUT-scale parameters, 
\beq
\left( \begin{array}{c}
M_Q^2 \\ M_U^2 \\ M_D^2 \\ M_L^2 \\ M_E^2 \\ M_G^2 \\ M_W^2 \\ M_B^2 
\end{array}
\right)
= 
\left( 
\begin{array}{cr}
1 & ~~6.97  \\
1 & 6.54  \\
1 & 6.49  \\
1 & 0.52  \\
1 & 0.15 \\
0 & 8.29     \\
0 & 0.68     \\
0 & 0.17
\end{array}
\right)
\left( 
\begin{array}{c}
M_{0}^2  \\ M_{1/2}^2 
\end{array}
\right).
\eeq
The last three equations express the gaugino masses in terms of a single parameter, $M_{1/2}$.
Noting the different proportionality coefficients, it is easy to see that the CMSSM predicts the 
gaugino mass hierarchy
\begin{equation}
M_G^2 > M_W^2 > M_B^2.
\label{eq:cmssm-gaugino-hierarchy}
\end{equation}
This observation immediately rules out the three 2-particle hierarchies 
\beq
\left\{BW,WG,BG\right\}.
\label{sugra2_3}
\eeq
Another way to arrive at this result is to note that one can eliminate the $M_{1/2}$ parameter
and obtain two independent gaugino mass sum rules, e.g.
\bea
0.68 M_G^2 - 8.29 M_W^2 &=&0, \label{GWsumrule} \\
0.17 M_W^2 - 0.68 M_B^2 &=&0, \label{WBsumrule}
\eea
which can be equivalently written as
\bea
-0.68 (M_W^2-M_G^2) - 7.61 M_W^2 &=&0, \label{GWsumrule-2} \\
-0.17 (M_B^2-M_W^2) - 0.51 M_B^2 &=&0. \label{WBsumrule-2}
\eea
Note that the numerical coefficients on the left-hand side of those equations are all negative.
Thus the hierarchy $WG$, where $M_W>M_G$, violates the first sum rule, eq.~(\ref{GWsumrule-2}),
while the hierarchy $BW$, where $M_B>M_W$, is in contradiction with the second sum 
rule, eq.~(\ref{WBsumrule-2}).\footnote{The impossibility of the third forbidden gaugino mass hierarchy, $BG$, follows 
from a linear combination of eqs.~(\ref{GWsumrule}) and (\ref{WBsumrule}), e.g.
$$
0.17 M_G^2 - 8.29 M_B^2 = -0.17 (M_B^2-M_G^2) - 8.12 M_B^2 = 0.
$$}

Similar logic can be applied to the sfermion sector, where the dependence on 
the parameter $M_0^2$ is the same for all sfermion masses squared, and the 
mass splittings are induced only due to the different dependence on $M_{1/2}^2$.
The CMSSM predicts the scalar masses in the order
\beq
M_Q^2>M_U^2>M_D^2>M_L^2>M_E^2,
\label{eq:cmssm-sfermion-hierarchy}
\eeq
which rules out ten additional 2-particle hierarchies, namely
\beq
\left\{ EL,ED,EU,EQ,LD,LU,LQ,DU,DQ,UQ \right\}.
\label{sugra2_10}
\eeq
This conclusion can also be justified with suitable sum rules, e.g. $EL$ is ruled out by the relation
\beq
-0.17 (M_E^2-M_L^2) - 0.37 M_B^2 = 0.
\eeq

Eqs.~(\ref{sugra2_3}) and (\ref{sugra2_10}) account for 13 of the 20 forbidden 2-particle hierarchies.
The remaining seven forbidden hierarchies, namely
\beq
\left\{ WQ,BQ,WU,BU,WD,BD,BL \right\}
\label{sugra2_7}
\eeq
can also be understood in terms of sum rules involving two sfermion masses and one gaugino mass:
\bea
WQ: \quad -6.82 (M_W^2-M_Q^2) - 6.14 M_Q^2 - 0.68 M_E^2 &=& 0,\\
BQ: \quad -6.82 (M_B^2-M_Q^2) - 6.65 M_Q^2 - 0.17 M_E^2 &=& 0, \\
WU: \quad -6.39 (M_W^2-M_U^2) - 5.71 M_U^2 - 0.68 M_E^2 &=& 0,\\
BU: \quad -6.39 (M_B^2-M_U^2) - 6.22 M_U^2 - 0.17 M_E^2 &=& 0, \\
WD: \quad -6.34 (M_W^2-M_D^2) - 5.66 M_D^2 - 0.68 M_E^2 &=& 0,\\
BD: \quad -6.34 (M_B^2-M_D^2) - 6.17 M_D^2 - 0.17 M_E^2 &=& 0, \\
BL: \quad -0.37 (M_B^2-M_L^2) - 0.20 M_L^2 - 0.17 M_E^2 &=& 0. 
\eea

As indicated in table~\ref{tab:forbidden}, in the CMSSM model the forbidden 2-particle hierarchies,
eqs.~(\ref{sugra2_3}), (\ref{sugra2_10}) and (\ref{sugra2_7}),
completely specify the forbidden $3$-particle hierarchies as well.  
However, there still remain twelve newly forbidden $4$-particle hierarchies 
which do not contain any forbidden $2$-particle hierarchies and need to be 
motivated by a different argument.

The idea is the following. Recall from Fig.~\ref{fig:case2} that in the 
allowed range for $\frac{M_{U}^2}{M_Q^2}$ there exist eight crossing points (\ref{defcross})
where we switch from a hierarchy $XY$ (to the left of the crossing point)
to a hierarchy $YX$ (to the right of the crossing point). Now consider two such crossing points,
$X_1Y_1\to Y_1X_1$ and $X_2Y_2\to Y_2X_2$, where all four particles $X_1$, $X_2$, $Y_1$ and $Y_2$ are different.
For definiteness let us also assume that
\beq
\left. \frac{M_{U}^2}{M_Q^2}\right|_{M_{X_1}=M_{Y_1}}  <
\left. \frac{M_{U}^2}{M_Q^2}\right|_{M_{X_2}=M_{Y_2}}. 
\label{crossassumption}
\eeq
Now consider the 4-particle hierarchy $Y_2X_2X_1Y_1$. The $X_1Y_1$ bit requires
\beq
\frac{M_{U}^2}{M_Q^2} < \left. \frac{M_{U}^2}{M_Q^2}\right|_{M_{X_1}=M_{Y_1}}, 
\label{X1Y1}
\eeq
while the $Y_2X_2$ bit in turn requires
\beq
\left. \frac{M_{U}^2}{M_Q^2}\right|_{M_{X_2}=M_{Y_2}} < \frac{M_{U}^2}{M_Q^2}. 
\label{X2Y2}
\eeq
In light of our original assumption (\ref{crossassumption}), eqs.~(\ref{X1Y1}) 
and (\ref{X2Y2}) place contradictory requirements on the parameter $\frac{M_{U}^2}{M_Q^2}$,
thus the hierarchy $Y_2X_2X_1Y_1$ is not allowed. 

From (\ref{intersectionpoints}) we find a total of 13 pairs of crossing points for which all 
four particles involved are different. This gives us 13 possible candidates for newly forbidden 4-particle hierarchies:
\begin{eqnarray}
&&\left\{QGBE,QGWL,QGWE,
UGBE,UGWL,UGWE, \right. \nonumber \\
&& \left. DGBE,DGWL,DGWE,
LGBE,LGWE,LWBE,EGLW
\right\}.
\label{eq:cmssm-other-twelve}
\end{eqnarray}
However, the last 4-particle hierarchy, $EGLW$, contains a forbidden 2-particle sub-hierarchy, $EL$,
and is therefore not newly forbidden. This leaves us with exactly 12 newly forbidden 4-particle hierarchies,
which are precisely those listed in (\ref{m1n4}).

\subsection{Examples of forbidden hierarchies in the most general case ($m=8$)}
\label{sec:sumrules}

In the previous subsection~(\ref{sec:CMSSMhierarchies}) we discussed the classification of forbidden 
hierarchies in a simple case like the CMSSM model. We shall now consider the other extreme, namely, the 
most general case of model $m=8$, where all 6 GUT scale input parameters are {\em a priori} unconstrained.
According to table~\ref{tab:forbidden}, in the case of model $m=8$, there are 32 forbidden 
4-particle hierarchies, which are listed in eq.~(\ref{m8n4}). In this subsection we shall illustrate a few specific examples.

As we saw in the case of the CMSSM, a mass hierarchy may be forbidden because it violates 
a mass sum rule. In the case of model $m=8$, there are 6 input model parameters, which enforces
two mass sum rules among the 8 observable mass parameters, e.g.
\bea
M_Q^2 + M_D^2 - M_L^2 - M_E^2 - 1.56 M_G^2 + 0.99 M_B^2 &=& 0 , \label{m8sr1} \\
M_U^2 + 2M_D^2 - 2 M_L^2 - M_E^2 - 2.34 M_G^2 +1.44 M_W^2 +0.74 M_B^2&=&0\, ,
\label{m8sr2}
\eea

\subsubsection{The mass hierarchy $GWQB$}

Let us first consider the forbidden 4-particle hierarchy $GWQB$
and rewrite the sum rules (\ref{m8sr1}) and (\ref{m8sr2}) in terms of the positive {\em mass differences}, e.g.
$M_G^2 -M_W^2$, $M_W^2-M_Q^2$ and $M_Q^2-M_B^2$:
\bea
-1.56(M_G^2-M_Q^2) 
-0.56 (M_Q^2-M_B^2)
+0.42 M_B^2 +M_D^2 -M_L^2 -M_E^2 &=& 0,  \label{GWQBsr1}\\
 -2.34 (M_G^2 -M_W^2) 
-0.91 (M_W^2-M_B^2)
-0.17 M_B^2 +M_U^2+2 M_D^2-2M_L^2-M_E^2&=&0.~~~~~~~  \label{GWQBsr2}
\eea
If this manipulation renders one of the sum rules (or a linear combination of them) 
in a form where all terms have numerical coefficients of the same sign, the 
hierarchy will be clearly forbidden, due to the sum rule. 
Unfortunately, this is not the case here --- the coefficients in both (\ref{GWQBsr1})
and (\ref{GWQBsr2}) have alternating signs, and the mass parameters can be suitably
adjusted to make the left-hand sides of those equations vanish. Therefore, the
sub-hierarchy $GWQB$ is not forbidden by the sum rules alone, and we need to
further investigate the GUT scale boundary conditions $\vec{G}^{8}$.
For this purpose, we shall invert eq.~(\ref{DeltaHNm}) and solve for
$\vec{G}^{8}$ in terms of the weak-scale mass parameters.
Because of the two sum rules (\ref{m8sr1}) and (\ref{m8sr2}), 
only 6 of the weak-scale mass-squared parameters are linearly independent, 
and one possible choice would be to supplement the masses of the 4 particles entering 
the hierarchy $GWQB$ under consideration, with the two slepton masses squared $M_L^2$ and $M_E^2$.
Then the relevant $d^{(8)}=6$ equations from the system (\ref{DeltaHNm}) are
\beq
 \begin{pmatrix}
M_G^2-M_W^2 \\
M_W^2-M_Q^2 \\
M_Q^2 - M_B^2 \\
M_B^2 \\
M_L^2 \\
M_E^2
\end{pmatrix}
= 
 \begin{pmatrix}
1 & -1 & 0 & 0 & 0 & 0 \\
0 &  1 & -1 & 0 & 0 & 0 \\
0 & 0 & 1 & -1 & 0 & 0 \\
0 & 0 & 0 & 1 & 0 & 0 \\
0 & 0 & 0 & 0 & 1 & 0 \\
0 & 0 & 0 & 0 & 0 & 1 \\
 \end{pmatrix}
 \begin{pmatrix}
0 & 0 & 0 & 0 & 0 & 1 & 0 & 0 \\
0 & 0 &  0& 0 & 0 & 0 & 1 & 0\\
1 & 0 &  0& 0 & 0 & 0 & 0 & 0\\
0 & 0 &  0& 0 & 0 & 0 & 0 & 1\\
0 & 0 &  0& 1 & 0 & 0 & 0 & 0\\
0 & 0 &  0& 0 & 1 & 0 & 0 & 0\\
 \end{pmatrix}
R^{(8)}
\left( 
\begin{array}{c}
M_{10}^2 \\ M_5^2 \\ \tilde S \\ M_{3}^2 \\ M_{2}^2 \\ M_{1}^2 
\end{array}
\right) ,
\label{eq:check8}
\eeq
Inverting this equation, we find
\beq
\left( 
\begin{array}{c}
M_{10}^2 \\ M_5^2 \\ \tilde S \\ M_{3}^2 \\ M_{2}^2 \\ M_{1}^2 
\end{array}
\right) 
=
\left(
\begin{array}{cccccc}
 -0.94 & -1.8 & -0.6 & -0.45 & 0 & -0.2 \\
 0.47 & 0.18 & -0.42 & -1.16 & 1 & 0.6 \\
 0.16 & 0.3 & 0.1 & -0.07 & 0 & 0.2 \\
 0.12 & 0.12 & 0.12 & 0.12 & 0 & 0 \\
 0 & 1.48 & 1.48 & 1.48 & 0 & 0 \\
 0 & 0 & 0 & 5.89 & 0 & 0 \\
\end{array}
\right)
 \begin{pmatrix}
M_G^2-M_W^2 \\
M_W^2-M_Q^2 \\
M_Q^2 - M_B^2 \\
M_B^2 \\
M_L^2 \\
M_E^2
\end{pmatrix},
\eeq
where we have substituted numeric values. Note, in particular, the first equation, which 
specifies the value of the GUT scale parameter $M_{10}^2$ as
\beq
M_{10}^2 =  -0.94(M_G^2-M_W^2)  -1.8(M_W^2-M_Q^2)  -0.6(M_Q^2 - M_B^2)  -0.45 M_B^2  -0.2M_E^2.
\eeq
Since all mass terms on the right-hand side have negative coefficients, 
the weak-scale masses squared are positive by definition, 
and the mass-squared differences are positive by the assumption 
of the hierarchy $GWQB$, 
we conclude that the $GWQB$ hierarchy necessarily requires a tachyonic value for 
$M_{10}^2$ and is therefore forbidden.
  
Recall that above we had the freedom of choosing two additional weak-scale parameters 
to form the vector in the left-hand side of (\ref{eq:check8}). It is instructive to see what happens 
if we had chosen a different set of weak-scale mass parameters, say $M_U^2$ and $M_L^2$ instead.
In that case eq.~(\ref{eq:check8}) is replaced with 
\beq
 \begin{pmatrix}
M_G^2-M_W^2 \\
M_W^2-M_Q^2 \\
M_Q^2 - M_B^2 \\
M_B^2 \\
M_U^2 \\
M_L^2
\end{pmatrix}
= 
 \begin{pmatrix*}[r]
1 & -1 & 0 & 0 & 0 & 0 \\
0 &  1 & -1 & 0 & 0 & 0 \\
0 & 0 & 1 & -1 & 0 & 0 \\
0 & 0 & 0 & 1 & 0 & 0 \\
0 & 0 & 0 & 0 & 1 & 0 \\
0 & 0 & 0 & 0 & 0 & 1 \\
 \end{pmatrix*}
 \begin{pmatrix}
0 & 0 & 0 & 0 & 0 & 1 & 0 & 0 \\
0 & 0 &  0& 0 & 0 & 0 & 1 & 0\\
1 & 0 &  0& 0 & 0 & 0 & 0 & 0\\
0 & 0 &  0& 0 & 0 & 0 & 0 & 1\\
0 & 1 &  0& 0 & 0 & 0 & 0 & 0\\
0 & 0 &  0& 1 & 0 & 0 & 0 & 0\\
 \end{pmatrix}
R^{(8)}
\left( 
\begin{array}{c}
M_{10}^2 \\ M_5^2 \\ \tilde S \\ M_{3}^2 \\ M_{2}^2 \\ M_{1}^2 
\end{array}
\right),
\label{eq:check8-2}
\eeq
and inverting, we get
\beq
\left( 
\begin{array}{c}
M_{10}^2 \\ M_5^2 \\ \tilde S \\ M_{3}^2 \\ M_{2}^2 \\ M_{1}^2 
\end{array}
\right)
=
\left(
\begin{array}{rrrrrr}
 -0.78 & -1.36 & -0.56 & -0.66 & 0.2 & 0 \\
 0 & -1.15 & -0.55 & -0.55 & -0.6 & 1 \\
 0 & -0.14 & 0.06 & 0.13 & -0.2 & 0 \\
 0.12 & 0.12 & 0.12 & 0.12 & 0 & 0 \\
 0 & 1.48 & 1.48 & 1.48 & 0 & 0 \\
 0 & 0 & 0 & 5.89 & 0 & 0 \\
\end{array}
\right)
 \begin{pmatrix}
M_G^2-M_W^2 \\
M_W^2-M_Q^2 \\
M_Q^2 - M_B^2 \\
M_B^2 \\
M_U^2 \\
M_L^2
\end{pmatrix}.
\label{orange}
\eeq
Note that the matrix on the right hand side does not have all negative entries in any given row,
so the contradiction is not immediately obvious. However, recall the existence of the two sum rules
(\ref{GWQBsr1}) and (\ref{GWQBsr2}), which define the remaining two parameters 
$M_D^2$ and $M_E^2$:
\beq
\left( 
\begin{array}{c}
M_{D}^2 \\ M_E^2  
\end{array}
\right)
=
\left(
\begin{array}{rrrrrr}
 0.78 & -0.66 & 0.34 & 0.59 & -1 & 1 \\
-0.78 & -2.22 & -0.22 & 1.02 & -1 & 0 \\
\end{array}
\right)
 \begin{pmatrix}
M_G^2-M_W^2 \\
M_W^2-M_Q^2 \\
M_Q^2 - M_B^2 \\
M_B^2 \\
M_U^2 \\
M_L^2 \\
\end{pmatrix}.
\label{blue}
\eeq
Putting together (\ref{orange}) and (\ref{blue}), we obtain
\beq
\left( 
\begin{array}{c}
M_{10}^2 \\ M_5^2 \\ \tilde S \\ M_{3}^2 \\ M_{2}^2 \\ M_{1}^2 \\ M_D^2 \\M_E^2
\end{array}
\right)
=
\left(
\begin{array}{rrrrrr}
 -0.78 & -1.36 & -0.56 & -0.66 & 0.2 & 0 \\
 0 & -1.15 & -0.55 & -0.55 & -0.6 & 1 \\
 0 & -0.14 & 0.06 & 0.13 & -0.2 & 0 \\
 0.12 & 0.12 & 0.12 & 0.12 & 0 & 0 \\
 0 & 1.48 & 1.48 & 1.48 & 0 & 0 \\
 0 & 0 & 0 & 5.89 & 0 & 0 \\
 0.78 & -0.66 & 0.34 & 0.59 & -1 & 1 \\
-0.78 & -2.22 & -0.22 & 1.02 & -1 & 0 \\ 
\end{array}
\right)
 \begin{pmatrix}
M_G^2-M_W^2 \\
M_W^2-M_Q^2 \\
M_Q^2 - M_B^2 \\
M_B^2 \\
M_U^2 \\
M_L^2
\end{pmatrix}.
\eeq
From a linear combination of the first and last equation, we obtain:
\bea
2M_{10}^2+M_E^2 &=&
-2.34(M_G^2-M_W^2) 
-4.93 (M_W^2-M_Q^2)  \nonumber \\
&-& 1.33 (M_Q^2 - M_B^2) 
- 0.3 M_B^2 
- 0.6 M_U^2 .
\eea
This sum rule cannot be satisfied because the LHS must be positive while the RHS is clearly 
negative. The lesson is that, depending on our choice of parameters for the 
inversion, we may have to use the sum rules (\ref{GWQBsr1}) and (\ref{GWQBsr2}) 
in order to show that a hierarchy is forbidden.

\subsubsection{The mass hierarchy $WQGL$}

As another example, consider the 4-particle sub-hierarchy $WQGL$.
The RGE solutions are
\beq
  \begin{pmatrix}
M_W^2-M_Q^2 \\
M_Q^2-M_G^2 \\
M_G^2 - M_L^2 \\
M_L^2 \\
M_U^2 \\
M_E^2
\end{pmatrix}
=
\left(
\begin{array}{rrrrrr}
 -1 &   0  & -1   & -6.48 &   0.19   & -0.004 \\
 1 &   0  &   1   & -1.81 &   0.49   &   0.004 \\
 0   & -1  &   3   &   8.29 & -0.49   & -0.04 \\
 0   &   1  & -3   &     0    &   0.49   &   0.04 \\
 1   &   0  & -4   &   6.48 &   0        &   0.07 \\
 1   &   0  &   6   &     0    &   0        &   0.15  \\
\end{array}
\right)
\left( 
\begin{array}{c}
M_{10}^2 \\ M_5^2 \\ \tilde S \\ M_{3}^2 \\ M_{2}^2 \\ M_{1}^2 
\end{array}
\right) 
\, .
\eeq
Selecting $M_U^2$ and $M_E^2$ as additional parameters to
invert the RGE equation, we have 
\beq
\left( 
\begin{array}{c}
M_{10}^2 \\ M_5^2 \\ \tilde S \\ M_{3}^2 \\ M_{2}^2 \\ M_{1}^2 
\end{array}
\right) 
=
\left(
\begin{array}{rrrrrr}
 -0.69 & 0.27 & -0.57 & -0.57 & 0.12 & -0.08 \\
 -1.15 & -0.55 & -0.55 & 0.45 & -0.6 & 0 \\
 -0.06 & 0.02 & 0.07 & 0.07 & -0.14 & 0.06 \\
 0 & 0 & 0.12 & 0.12 & 0 & 0 \\
 1.48 & 1.48 & 1.48 & 1.48 & 0 & 0 \\
 6.86 & -2.68 & 1.05 & 1.05 & 4.77 & 4.77 \\
\end{array}
\right)
  \begin{pmatrix}
M_W^2-M_Q^2 \\
M_Q^2-M_G^2 \\
M_G^2 - M_L^2 \\
M_L^2 \\
M_U^2 \\
M_E^2
\end{pmatrix} \, .
\eeq
Again, we do not have a single row with negative coefficients, but
forming the sum of $M_{10}^2$ and $M_5^2$ we find
\bea
M_{10}^2 + M_5^2 = &&
-1.84 \left(M_W^2-M_Q^2\right)-0.28\left(M_Q^2-M_G^2\right) \nonumber \\
&& -1.12\left(M_G^2 - M_L^2\right)-0.12\, M_L^2 -0.48 \, M_U^2-0.08\, M_E^2 .
\eea
This sum rule cannot be satisfied, since all terms in the RHS 
have negative coefficients. Therefore, the hierarchy $WQGL$ is forbidden.

\section{Conclusions and summary}
\label{sec:conclusions}

The search for SUSY is the paramount experimental challenge for Run II
at the LHC.  Once SUSY is discovered, it may provide invaluable clues
about GUT-scale physics. The measured pattern of SUSY particle masses  
will play an important role in this quest, as we have explained above.
Our proposal is to consider the relative ordering, or the ``hierarchy", 
of the measured sparticle masses. By analyzing the one loop SUSY RGE's,
it is relatively straightforward to derive the complete set of allowed hierarchies 
for a given choice of GUT scale boundary conditions. In this paper we 
considered hierarchies involving up to eight weak scale masses
(first/second generation masses for each of the five sfermion
families, and three gaugino masses), and analyzed eight different 
GUT-scale model scenarios (table~\ref{tab:models}).
Our results are listed in appendix~\ref{app:forbidden}
and can be reproduced with the accompanying {\tt Python} code
described in appendix~\ref{sec:python-code}.
We also provided some intuitive arguments, based on mass sum rules and 
linear algebra tricks, to better understand and justify those results.

\begin{figure}[t]
\begin{center}
\includegraphics[width=0.8\textwidth]{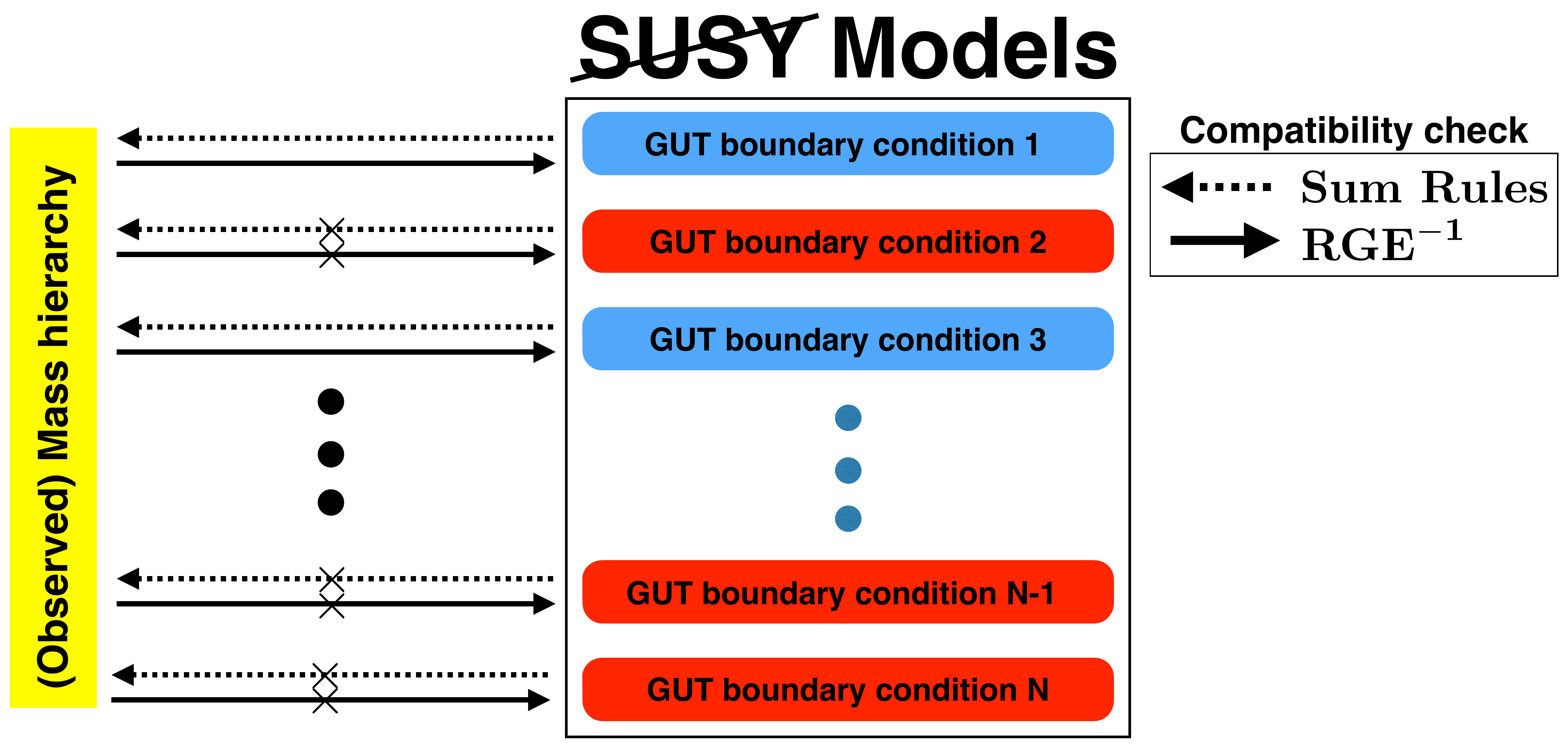} 
\end{center}
\caption{ A schematic view on how hierarchies shed light on GUT-scale boundary conditions.
Once a set of SUSY particle masses is observed (left), GUT-scale boundary conditions (center) incapable 
of producing this sparticle mass hierarchy are ruled out by querying our database.  
This conclusion can be independently verified by 
suitable mass sum rules or explicit running of the RGEs (right).
\label{fig:map}}
\end{figure}

The advantage of the approach presented in this paper is that it allows to draw
definitive conclusions based on only partial information. In particular, we have seen that 
the knowledge of a {\em sub-hierarchy} of a few sparticle masses, measured in the 
very early days of discovery, is sometimes sufficient to rule out a specific model.
Consider, for example, the most general model, $m=8$, which has 6 GUT-scale input parameters.
In principle, one would need 6 independent measurements in order to fully 
reconstruct the GUT-scale physics. On the other hand, we have shown that 
in this model, there are as many as 32 forbidden $4$-particle sub-hierarchies.
Therefore, the $m=8$ model can be ruled out with only 4 suitable measurements,
if the data happens to point to one of the 32 forbidden hierarchies.
This procedure is pictorially illustrated in figure~\ref{fig:map}.

The analysis presented here can be extended in many directions. 
Obviously, one may consider more of the MSSM particles, or 
generalize to non-minimal models like the NMSSM. 
Our approach is not limited to SUGRA-type boundary conditions and 
can be applied in other SUSY breaking scenarios as well.
Armed with the toolkit of allowed and forbidden hierarchies, 
we look forward to analyzing GUT scale models after a SUSY discovery
at the LHC!

\acknowledgments
We thank P.~Konar and G.~Sarangi for collaboration in the initial
stages of this work and A.~de Gouvea and I.~Low for useful comments
and suggestions.  MP thanks the high energy theory group at the University of
Florida for its warm and humid hospitality.  
MP is supported by the Korea Ministry of Science, ICT and Future Planning, 
Gyeongsangbuk-Do and Pohang City for Independent Junior Research Groups 
at the Asia Pacific Center for Theoretical Physics.
MP is also supported by World Premier International Research Center 
Initiative (WPI Initiative), MEXT, Japan.
Work also supported by DOE Grant
No.~DE-SC0010296.

\appendix

\section{Results on forbidden hierarchies}
\label{app:forbidden}

In this appendix, we provide explicit lists of ``irreducibly-forbidden'' or
``newly-forbidden'' $n$-particle sub-hierarchies in various GUT scale
scenarios (``models'')\footnote{For the definition of the eight models, see table~\ref{tab:models}.}.
These are the forbidden $n$-particle
hierarchies which do not contain a forbidden sub-hierarchy.

\subsection{Model $m=1$ (CMSSM)}

The forbidden (and hence irreducibly-forbidden) 2-particle hierarchies are
\bea
&& WU,
BW,
DQ,
BL,
ED,
BU,
WQ,
LQ,
EQ,
EL, \nonumber \\
&&
LD,
DU,
WD,
BG,
WG,
BQ,
LU,
EU,
BD,
UQ.
\label{m1n2}
\eea
The irreducibly-forbidden 4-particle hierarchies are
\bea
&& UGBE,
QGWE,
UGWL,
LWBE,
QGWL,
DGBE, \nonumber \\
&&
DGWL,
QGBE,
LGWE,
LGBE,
UGWE,
DGWE.
\label{m1n4}
\eea
The are no newly forbidden $n$-particle hierarchies for $n\ge5$ in this model.
It is astonishing that all the information about the $8!$ possible hierarchies in the CMSSM is encoded in
the two lists (\ref{m1n2}) and (\ref{m1n2}).

\subsection{Model $m=2$}

For model $2$, the forbidden 2-particle hierarchies are
\bea
WU,BW,BL,BU,WQ,EQ,LD,WD,BG,WG,BQ,EU,BD,UQ.
\label{m2n2}
\eea
The newly forbidden 4-particle hierarchies are
\bea
&& GDLU,UGBE,DQWL,QGWE,GDLQ,GUED,DUEL,  \nonumber \\
&& DGWL,QGBE,UGWE,QDLU,DQEL,GQED;
\label{m2n4}
\eea
and the newly forbidden 6-particle hierarchies are
\bea
&& UDLWBE,QGDUWL.
\label{m2n6}
\eea
There are no other newly forbidden hierarchies.

\subsection{Model $m=3$ (NUHM)}

For this model, the forbidden 2-particle hierarchies are
\bea
BW,WQ,WD,BG,WG,BQ,BD;
\label{m3n2}
\eea
the newly-forbidden 3-particle hierarchies are
\bea
&&DUQ,UEQ,EQD,LUQ,UDQ,LEU,GDQ,EUD,ELQ,LUD,ELD,LQU, \nonumber \\
&&LEQ,EUQ,WEU,DGQ,LED,LUE,BEU,LGQ,DLQ,WUE,EGQ,LGD, \nonumber \\
&&EGD,QED,LDQ,BUE,GLD,GED,UED,LDU,BLE,GLQ,BEL,GEQ;
\label{m3n3}
\eea
the newly forbidden 4-particle hierarchies are
\bea
&&DQBE,GDLU,UGBL,DEBU,UQWL,DUBE,QDEU,DELU,GQLU, \nonumber \\
&&GDWU,GQBU,GDEU,UGWL,DEWU,QEWU,WEBL,GEBU,QELU, \nonumber \\
&&GQEU,GDBU,GQWU,GELU,QEBU,UDBL,QDWU,UDEL,UQEL, \nonumber \\
&&QDLU,GEWU,QULD,DQWE,QDBU,UQBL,DQLE;
\label{m3n4}
\eea
the newly forbidden 5-particle hierarchies are
\bea
&&QGWEL,QGUWE,DGWEL,QGUBE,QGWLE,\nonumber \\
&&DGUWE,QUGWE,DUGWE,DGWLE,QUGBE; \label{m3n5}
\eea
and the newly forbidden 6-particle hierarchies are
\bea
QGUDWL,QDEGWL,QDEGBL.
\label{m3n6}
\eea
There are no newly forbidden $7$-particle or $8$-particle hierarchies in this model.

\subsection{Model $m=4$}

For this model, the forbidden 2-particle hierarchies
are
\bea
BW,WQ,BG,WG,BQ;
\label{m4n2}
\eea
and the newly-forbidden 3-particle hierarchies are
\bea
&&UEQ,WDU,BLD,BUD,EUQ,WEU,BEU,WUE,EGQ, \nonumber \\
&&BUE,WLD,BDU,BLE,WUD,BEL,BDL,WDL,GEQ. 
\label{m4n3}
\eea
Additionally, as shown in table~\ref{tab:forbidden}, there are 
79 newly forbidden $4$-particle hierarchies, 
14 newly forbidden $5$-particle hierarchies, and
3 newly forbidden $6$-particle hierarchies.
(There are no newly forbidden $7$-particle or $8$-particle hierarchies.)
In order to save space, we do not list these hierarchies explicitly here,
however they can be easily obtained from the python code described in
appendix~\ref{sec:python-code}.



\subsection{Model $m=5$}

There is only one forbidden 2-particle hierarchy
\bea
DU,
\label{m5n2}
\eea
and six newly-forbidden 3-particle hierarchies:
\bea
BGU,GWQ,WGQ,EGU,GBU,GEU.
\label{m5n3}
\eea
Additionally, as shown in table~\ref{tab:forbidden}, there are 
121 newly forbidden $4$-particle hierarchies, 
168 newly forbidden $5$-particle hierarchies,
149 newly forbidden $6$-particle hierarchies, and 
12 newly forbidden $7$-particle hierarchies.
(There are no newly forbidden $8$-particle hierarchies.)
In order to save space, we do not list these hierarchies explicitly here,
however they can be easily obtained from the python code described in
appendix~\ref{sec:python-code}.




\subsection{Model $m=6$}

In this model, all 2-particle hierarchies are allowed. The forbidden
3-particle hierarchies are:
\bea
BGU,GWQ,EGU,GBU,GEU,WGQ.
\label{m6n3}
\eea
The newly-forbidden 4-particle hierarchies are 
\bea
&&UEWQ,WEUQ,WDLQ,LQGD,WBUQ,LGQD,UQWE,WQDL,WULQ, \nonumber \\
&&UQWL,DQWL,UWQE,GDLQ,WQUL,UWEQ,GUED,WUQL,UBWQ, \nonumber \\
&&WDQL,UWQB,GUBQ,LGDQ,GQWU,WUEQ,GQWD,BUGE,UQWB,\nonumber \\
&&WUQB,DLWQ,DWQL,DWLQ,GLQD,EUGB,WUQE,UWBQ,UWLQ, \nonumber \\
&&WUBQ,WLDQ,UWQL,GUEQ,GUBD,WQGL,GQLD,GLDQ,WLUQ. 
\label{m6n4}
\eea
Additionally, as shown in table~\ref{tab:forbidden}, there are 
216 newly forbidden $5$-particle hierarchies, 
288 newly forbidden $6$-particle hierarchies,
98 newly forbidden $7$-particle hierarchies, and 
3 newly forbidden $8$-particle hierarchies.
In order to save space, we do not list these hierarchies explicitly here,
however they can be easily obtained from the python code described in
appendix~\ref{sec:python-code}.



\subsection{Model $m=7$}

In model $7$, all 2-particle and 3-particle hierarchies are allowed.
The forbidden 4-particle hierarchies are:
\bea
&& LQGU,WDQU,BGUE,GWQB,GULQ,WDEQ,DLQU,DWQB,DULQ,WDLQ, \nonumber \\
&& LUGQ,DBEU,WGQU,DEBU,GLWQ,WULQ,LGUQ,UQWL,WDBQ,DWUQ, \nonumber \\
&& WGLQ,DUBE,DBWQ,WQUL,GQLU,BEDU,WUGQ,LDEQ,GBUE,BGEU, \nonumber \\
&& BDUE,WGQB,GBWQ,WUQL,BEGU,WDQL,GEBU,GBEU,WDQB,GWBQ, \nonumber \\
&& GQWU,LUDQ,DEWQ,GUWQ,DLEQ,DELQ,DWQE,DWEQ,GWUQ,BUGE, \nonumber \\
&& GWQL,WEDQ,WUDQ,DLWQ,WGQL,LEDQ,DWQL,DUWQ,DWLQ,WDQE, \nonumber \\
&& DBUE,LGQU,WDUQ,DLQE,WBGQ,WLGQ,BDEU,LDUQ,BUDE,LQDU, \nonumber \\
&& DQWE,UWLQ,WGUQ,LDQU,LDQE,WLDQ,DLUQ,WBDQ,DWQU,WGBQ, \nonumber \\
&& UWQL,GWQU,DQLE,GWLQ,WQGL,GLUQ,DWBQ,GLQU,WLUQ.
\label{m7n4}
\eea
Additionally, as shown in table~\ref{tab:forbidden}, there are 
176 newly forbidden $5$-particle hierarchies, 
426 newly forbidden $6$-particle hierarchies,
434 newly forbidden $7$-particle hierarchies, and 
22 newly forbidden $8$-particle hierarchies.
In order to save space, we do not list these hierarchies explicitly here,
however they can be easily obtained from the python code described in
appendix~\ref{sec:python-code}.



\subsection{Model $m=8$}

In this model, all 2-particle and 3-particle hierarchies are allowed.
The forbidden 4-particle hierarchies are:
\bea
&& BGEU,GBWQ,WUQL,GEBU,GWQL,WBGQ,WGBQ,WGLQ, \nonumber \\
&& BEGU,WGQL,GBEU,UWQL,BGUE,GWQB,WUGQ,GUWQ, \nonumber \\
&& UWLQ,WGQB,WGUQ,WLUQ,UQWL,WGQU,GWBQ,GWUQ, \nonumber \\
&& GBUE,GWQU,GWLQ,WULQ,GLWQ,WQGL,WLGQ,WQUL.
\label{m8n4}
\eea
Additionally, as shown in table~\ref{tab:forbidden}, there are 
148 newly forbidden $5$-particle hierarchies, 
809 newly forbidden $6$-particle hierarchies,
398 newly forbidden $7$-particle hierarchies, and 
54 newly forbidden $8$-particle hierarchies.
In order to save space, we do not list these hierarchies explicitly here,
however they can be easily obtained from the python code described in
appendix~\ref{sec:python-code}.

\section{Python package for studying hierarchies}
\label{sec:python-code}

We have provided a {\tt Python}~\cite{Python} package in the
supplemental material to the arXiv version of this paper.
It consists of two required files, {\tt susy\_hierarchy\_methods.py}, and 
{\tt allowed\_hierarchies.py} and two optional files, {\tt
  forbidden\_hierarchies.py} and \\ {\tt
  newly\_forbidden\_hierarchies.py}.  These files should work with all
versions of {\tt Python} including and subsequent to {\tt Python 2.6}.

Information about which hierarchies are allowed or forbidden in
various SUSY scenarios is contained in the files~{\tt allowed\_hierarchies.py} and~{\tt
  forbidden\_hierarchies.py}.  Of course a hierarchy in a given SUSY
scenario is either allowed or forbidden, so it is sufficient to have
only, e.g., an enumeration of allowed hierarchies in various models.
However, the generation of forbidden hierarchy information from
allowed hierarchy information is relatively slow $\sim 1$ minute on a
modern laptop, so we do include the, in principle redundant, files {\tt
  forbidden\_hierarchies.py} and {\tt newly\_forbidden\_hierarchies.py}.

\subsection{Data structures in Python} 

To explain how hierarchy data is stored in the files {\tt
allowed\_hierarchies.py}, \\ {\tt
  forbidden\_hierarchies.py}, and {\tt
  newly\_forbidden\_hierarchies.py}, we must briefly review a few of
the data
structures present in Python.  A \textbf{list} consists of elements,
in a fixed order, between the square brackets `[' and `['.
Elements are indexed by integers, e.g., the first element of {\tt
  my\_list} is {\tt my\_list[0]}.  A \textbf{dictionary} is like a list, but
the elements are stored between curly brackets `\{' and `\}' and
are indexed by arbitrary items called ``keys''.  An element of a
dictionary is accessed by its key.  For example if 
\begin{verbatim}
my_dict = {'example':37, 32.3:'different'}
\end{verbatim}
then
\begin{verbatim}
my_dict['example'] = 37
my_dict['32.3'] = 'different'
\end{verbatim}
Finally a \textbf{tuple} is like a list, except that tuples cannot be
modified (and tuples are enclosed in parentheses).

\subsection{Hierarchy Data}

The file {\tt allowed\_hierarchies.py} contains the single command
\begin{verbatim}
allowed = {2 : {('L', 'W'): [1, 3, 5, 2, 7, 4, 6, 8], ('Q','G'): \
[1, 3, 5, 2, 7, 4, 6, 8], ....
\end{verbatim}
The dictionary ``allowed'' has keys $2$, $3$, ..., $8$; these keys
represent the number of particles in a hierarchy.  The elements
corresponding to each of these keys are dictionaries.  The keys for
these dictionary are the hierarchies, which are tuples of capital
letters representing the particles in the hierarchy.  The entry
corresponding to a hierarchy is a list (in no particular order) of the
SUSY scenarios in which the hierarchy is allowed.  The numbering of
SUSY scenarios or ``cases'' is that which is used throughout this
paper and defined in table~\ref{tab:models}.  
Note that only hierarchies which are allowed in \textit{some}
hierarchy have entries in ``allowed''.  Thus
\begin{verbatim}
print(allowed[2][('U','Q')])
\end{verbatim}
yields
\begin{verbatim}
[3, 5, 7, 4, 6, 8]
\end{verbatim}
as the 2-particle hierarchy \verb^('U', 'Q')^ is allowed in models $3$,
$4$, $5$, $6$, $7$, and $8$, while
\begin{verbatim}
print(allowed[4][('B', 'G', 'E', 'U')])
\end{verbatim}
yields the error (in interactive session):
\begin{verbatim}
Traceback (most recent call last):
File "<stdin>", line 1, in <module>
KeyError: ('B', 'G', 'E', 'U')
\end{verbatim}
as the 4-particle hierarchy \verb^('B', 'G', 'E', 'U')^ is not allowed
in any model.

The file {\tt forbidden\_hierarchies.py} uses the exact same format,
but the dictionary is called ``forbidden'' and contains lists of the
models in which hierarchies are forbidden, rather than allowed.
Likewise the file {\tt newly\_forbidden\_hierarchies.py} uses a
dictionary called \verb^``newly_forbidden''^, which tells us which
hierarchies are newly forbidden (irreducible).

\subsection{Methods}

The {\tt Python} expert can use the data in {\tt
  allowed\_hierarchies.py}, {\tt forbidden\_hierarchies.py}, and {\tt
  newly\_forbidden\_hierarchies.py} for
their own purposes.  However, for those less familiar with {\tt
  Python} or for those who would benefit from an example of methods
(functions) which use the hierarchy data, we have provided another
module (python file) with methods that process the data in {\tt
  allowed\_hierarchies.py}, {\tt forbidden\_hierarchies.py}, and {\tt
  newly\_forbidden\_hierarchies.py}.  These methods are 
\begin{enumerate}
\item {\tt convert\_to\_standard\_form(s)} \\
This method translates a string, $s$, to a tuple of capital letters
representing the SUSY particles in a hierarchy.  This is the format of
the keys representing hierarchies in our data arrays.
\item {\tt print\_case(case)} \\
This method prints a string describing the SUSY scenario (``case'')
labelled by the integer $case$, which must be between one and eight (inclusive).
\item {\tt get\_info\_on\_hierarchy(s) } \\
This method prints information listing and describing the SUSY scenarios
(``cases'') in which the hierarchy $s$ ($s$ is a string, enclosed by
single or double quotes) is allowed and in which it is forbidden.
\item {\tt number\_of\_N\_hierarchies\_in\_case(data\_type,N,case) } \\
This method returns the number of $N$ particle hierarchies which are
either allowed or forbidden (depending on whether the value of
$data\_type$ is ``allowed'' or ``forbidden'') in SUSY scenario $case$.
\item {\tt list\_of\_N\_hierarchies\_in\_case(data\_type,N,case) } \\
This method returns a list of $N$ particle hierarchies which are
either allowed or forbidden (depending on whether the value of
$data\_type$ is ``allowed'' or ``forbidden'') in SUSY scenario $case$.
\item {\tt print\_table\_of\_cases() } \\
This method prints a table that describes the eight SUSY scenarios
(``cases'') that we consider.
\item {\tt print\_table\_of\_number\_of\_hierarchies(data\_type) } \\
This method prints a table of all particle hierarchies which are
either allowed or forbidden (depending on whether the value of
$data\_type$ is ``allowed'' or ``forbidden'') in all eight SUSY
scenarios.
\end{enumerate}

\subsection{Using the code}

The {\tt Python} expert can use the methods and as part of their own
code by importing whichever modules they need.  However, the simplest
way to use the methods provided is to use {\tt Python} in interactive
mode.  To do this, one types
\begin{verbatim}
python
\end{verbatim}
in a terminal.  While it is discouraged in general (as it negates
useful features of {\tt Python} in regard to namespaces), the simplest
next step is to type
\begin{verbatim}
from susy_hierarchy_methods import *
\end{verbatim}
into the {\tt Python} interpreter.  Note that the files {\tt
  susy\_hierarchy\_methods.py} and \\ {\tt allowed\_hierarchies.py} must be
in the current working directory, or must be in one's {\tt PYTHONPATH}.
Doing this also imports ``allowed'' from {\tt allowed\_hierarchies.py}
automatically.  If {\tt forbidden\_hierarchies.py} is in the working
directory or the {\tt PYTHONPATH}, then ``forbidden'' will be imported
from this file.  If not, ``forbidden'' will be generated automatically,
though this can be somewhat time-consuming ($\sim 1$ minute on a
modern laptop).  The situation regarding {\tt
  newly\_forbidden\_hierarchies.py} is exactly analogous.
One can then simply type the method one wishes to execute together
with the appropriate arguments.  Finally, we note that typing
\begin{verbatim}
help(name_of_method)
\end{verbatim}
will provide a brief description of the method in question.

\end{document}